\documentclass[12p]{article}
\usepackage{latexsym}
\usepackage{color}
\usepackage{amsmath}
\usepackage{epsfig}
\usepackage{hyperref}
 \usepackage{dcolumn}
   \usepackage{threeparttable}

\newcommand{\ortala}[1]{\begin{center}#1\end{center}}

\newcommand{\sandd}[1]{\left\langle #1\right\rangle}

\newcommand{\intego}[3]{{{\underset{#1
}{\overset{#2}{\displaystyle\oint}}}#3}}
\newcommand{\summ}[3]{{{\underset{#1 }{\overset{#2}{\displaystyle\sum}}}#3}}

\newcommand{\re}[1]{(\ref{#1})}

\newcommand{\eq}[2]{\begin{equation}\label{#1}  #2\end{equation}}
\newcommand{\tur}[2]{\frac{d#1}{d#2}}

\newcommand{\paran}[1]{\left(#1\right)}

\newcommand{\sch}[1]{Schrodinger}

\newcommand{\komb}[2]{\paran{\begin{array}{c} #1 \\ #2 \end{array}}}

\linethickness{0.6pt}

\begin{document}

\ortala{\textbf{Hysteresis behavior of the anisotropic quantum Heisenberg model driven by periodic magnetic field}}

\ortala{\textbf{\"Umit Ak\i nc\i \footnote{umit.akinci@deu.edu.tr}}}

\ortala{\textit{Department of Physics, Dokuz Eyl\"ul University,
TR-35160 Izmir, Turkey}}

\section{Abstract}

Dynamic behavior of a quantum Heisenberg ferromagnet in the presence of a periodically
oscillating magnetic field has been analyzed by means of the effective field theory with two spin cluster.
The dynamic equation of motion has been constructed with the
help of a Glauber type stochastic process and solved for a simple cubic  lattice.
After the phase diagrams given, the behavior of the hysteresis loop area, coercive field
and remanent magnetization with the anisotropy in the exchange interaction
has been investigated in detail. Especially, by comparing of the magnitudes of
the hysteresis loop area in the high anisotropy limit (i.e. Ising model)
and low anisotropy limit (i.e. isotropic Heisenberg model), detailed
description of the hysteresis loop area with the anisotropy in the
exchange interaction given. Some interesting features have been obtained
about this behavior as well as in phase diagrams such as tricritical points.

Keywords: \textbf{Dynamic quantum anisotropic Heisenberg model;
hysteresis loops; hysteresis loop area; coercive field; remanent magnetization}

\section{Introduction}\label{introduction}

Recently, magnetic systems under a time dependent external magnetic field has been attracted
much interest from both theoretical and experimental points of view. It is a fact
that the competition between the time scales of the relaxation
time of the magnetically interacting system
and  period of the external applied magnetic field causes the
observations of unusual and interesting dynamic behaviors. If the relaxation time
is less than the period of the magnetic field, the time dependent magnetization of
the system can follow the magnetic field. In contrary to this,
when the frequency of the magnetic field rises then  after a specific value of
frequency, which sensitively depends on the amplitude of the magnetic field and
the temperature as well as the geometry of the lattice and exchange interaction,
it is able to follow the external magnetic field with phase lag. The physical mechanism
discussed briefly above indicates the existence of a pure
dynamic phase transition (DPT) \cite{ref1}. The time average of the magnetization
over a full period of the oscillating magnetic field can be used as
dynamic order parameter (DOP) of the system.

It is possible to mention that the another interesting behavior is hysteresis behavior,
which is a common behavior and occurs in  most of the physical systems. It originates
from the delay of the response of the system to the driving cyclic force.
Magnetic hysteresis (which is simply the
variation of the magnetization with  the driving magnetic field)
is one of the most important  features of the magnetic materials.
By benefiting from two important fundamental tools,
namely coercive field (CF) and remanence magnetization (RM),
one can make prediction on the shape of the hysteresis loops.
CF is defined as the intensity of the external magnetic field needed
to change the sign of the magnetization.  RM is residual
magnetization of a ferromagnetic material after an external
magnetic field is removed. In addition to these, there exists an
important quantity which is hysteresis loop area (HLA) corresponding
to the energy loos due to the hysteretic behavior.

Due to recent developments experimental techniques,
DPTs and hysteresis behaviors can be observed
experimentally in different types of magnetic systems.
Experiments on ultrathin Co films \cite{ref2},  Fe/Au(001)
films \cite{ref3}, epitaxial Fe/GaAs(001) thin
films \cite{ref4},  fcc Co(001), and fcc NiFe/Cu/Co(001) layers \cite{ref5}
Fe/InAs(001) ultrathin films \cite{ref6} are among them.

From  the theoretical point of view, DPT is first observed within the mean field
approximation calculations (MFA) \cite{ref7} for the s-$1/2$ Ising model.  Since than,
DPT and hysteresis behaviors of the s-$1/2$ Ising model have been widely studied within the
several techniques such as MFA \cite{ref8}, Monte Carlo simulation (MC) \cite{ref9},
effective field theory (EFT) \cite{ref10}. Besides, in the Ising case systems
with higher spins \cite{ref11} and mixed spins \cite{ref12} have been studied. One
can clearly think that Ising model can be viewed as highly
anisotropic case of the Heisenberg model. Actually,
since all of the magnetic materials do not exhibit this type of anisotropy,
it is necessary to consider Heisenberg model to provide more
realistic description of magnetic systems. In this way, it is obvious that a
much more physical information can be obtained within the
quantum Heisenberg model which includes the quantum
fluctuations.

In the Heisenberg scheme, DPT and hysteresis behaviors mostly have been investigated
in the classical case within the MC.  Classical isotropic  Heisenberg model
\cite{ref13,ref14,ref15} and uniaxially classical anisotropic  Heisenberg
model driven by sinusoidal magnetic field   \cite{ref16,ref17,ref19,ref20,ref21,ref22,ref23,ref24}
have been studied within the MC.  In the case of dilute Heisenberg model \cite{ref25},
classical anisotropic Heisenberg model on thin film geometry  \cite{ref26,ref27,ref28,ref29}
have been studied within the MC also. Besides, classical anisotropic Heisenberg model
within the  MFA \cite{ref30} and  classical anisotropic Heisenberg model on thin film
geometry in comparison with MFA and MC \cite{ref31}  have been studied. The readers may refer to
\cite{ref32} for a detailed discussion of these types of systems.

It can be easily seen from literature mentioned briefly above that MC simulation is
widely used to determine the true dynamic critic nature of these types of
magnetic systems. However, from the computational investigation
point of view,  MC simulations for the quantum cases are expensive choice than
the classical one. Hence, in order to overcome this difficulty, some other
well defined approximate techniques required. Of course, one of the methods
coming to mind is MFA. But, it is a well known fact that, in MFA,
self spin correlations are ignored and the results to be obtained do
not reflect the real details of system. Because EFT takes into account
the self spin correlations, it gives results that are superior to
those of obtained within the MFA. From this point of view, the aim of this work
is to investigate the anisotropic quantum Heisenberg model under a magnetic field
oscillating in time within the EFT formulation. For this aim the paper is organized
as follows: In Sec. \ref{formulation} we briefly present the
model and  formulation. The results and discussions are
presented in Sec. \ref{results}, and finally Sec. \ref{conclusion} contains our conclusions.

\section{Model and Formulation}\label{formulation}

We consider a lattice consisting  of $N$ identical spins of (s-$1/2$) such that each of
the spins has $z$ nearest neighbors. The Hamiltonian of the kinetic Heisenberg model is
given by \eq{denk1}{\mathcal{H}=-\summ{<i,j>}{}{\paran{J_x s_i^xs_j^x+J_y s_i^ys_j^y+J_z s_i^zs_j^z}}-H(t)\summ{i}{}{s_i^z}}
where $s_i^x,s_i^y$ and  $s_i^z$ denote the Pauli spin operators at a site $i$, $J_x,J_y$ and $J_z$
stand for the anisotropy in the exchange interactions between the nearest neighbor
spins and $H(t)$ is the time dependent external longitudinal magnetic field, respectively.
The first sum is over the nearest neighbors of the lattice, while the
second one is over all the lattice sites. Time dependent magnetic
field defined as \eq{denk2}{H(t)=H_0\sin{\paran{\omega t}}}
where $H_0$ is the amplitude and $\omega$ is the angular frequency of
the periodic magnetic field.

We use EFT-2 (two spin cluster EFT) formulation \cite{ref33} here,
which based on using generalized versions of the Callen-Suzuki spin identities \cite{ref34,ref35}
on the two spin clusters \cite{ref36}. These identities can be expanded with using
differential operator technique \cite{ref37}. When one expands these identities with differential
operator technique, multi spin correlations appear and in order to avoid from the mathematical
difficulties, these multi spin correlations often neglected by using decoupling
approximation \cite{ref38}. In EFT-2 formulation, interaction between the
chosen two spins  (namely $s_1$ and $s_2$) treated exactly. These two spins
constitute two spin cluster. The interaction between this cluster and
outside of it treated approximately. In order to avoid some mathematical
difficulties replacing of the perimeter spins of the two spin cluster by
Ising spins (axial approximation), is typical \cite{ref39}.  With using
translational invariancy of the lattice, defined variable $m=\sandd{s_i^z}=\sandd{\frac{1}{2}\paran{s_1^z+s_2^z}}$
can be used as magnetization per site. If one uses a Glauber-type stochastic process to
investigate dynamic properties of the considered system \cite{ref40}, one can obtain
dynamic equation of motion within the EFT-2 formulation as
\eq{denk3}{\theta \tur{\sandd{s_i^z}}{t}=-\sandd{s_i^z}+\sandd{\frac{a_1+
a_2}{X}\frac{\sinh{\paran{\beta X}}}{\cosh{\paran{\beta X}}+\exp{\paran{-2\beta J_z}}\cosh{\paran{\beta Y}}}}
,} where  $\theta$ is the transition rate per
unit time, $\beta=1/(k_BT)$ and $k_B$ and $T$ denote the Boltzmann constant and
temperature, respectively. The terms $X$ and $Y$ given by \eq{denk4}{
X=\left[\paran{J_x-J_y}^2+(a_1+a_2)^2\right]^{1/2}, \quad Y=\left[\paran{J_x+J_y}^2+(a_1-a_2)^2\right]^{1/2}
,}
where $a_1,a_2$ stand for the the local field acting on the   lattice sites $1$ and $2$ in chosen
cluster, respectively.  These local fields include the interaction of the spins in
chosen cluster with the nearest neighbor spins belongs to the outside of the cluster and magnetic field,
\eq{denk5}{
a_i=J_z\summ{\delta}{}{s_{i,\delta}^z}+H_i,
} where $s_{i,\delta}^z$ denotes the nearest neighbor of the spin $s_i^z$ and $H_i$
is external periodic magnetic field at a site $i$ ($i=1,2$).

Thermal average of the right hand side of Eq. \re{denk3} can be
handled by  differential operator technique and decoupling approximation (DA) \cite{ref38} as

\eq{denk6}{
\sandd{\frac{a_1+a_2}{X_0}\frac{\sinh{\paran{\beta X_0}}}{\cosh{\paran{\beta X_0}}+\exp{\paran{-2\beta J_z}}\cosh{\paran{\beta Y_0}}}}
=\sandd{\left[A_{x}+m B_{x}\right]^{z_0}
\left[A_{y}+m B_{y}\right]^{z_0}
\left[A_{xy}+m B_{xy}\right]^{z_1}} f\paran{x,y,H_1,H_2}|_{x=0,y=0}
} where each of $s_1$ and $s_2$ has number of $z_0$ distinct nearest neighbors and both of them have $z_1$ common nearest neighbor.
The function in Eq. \re{denk6} is given by
\eq{denk7}{f\paran{x,y,H_1,H_2}=\frac{x+y+H_1+H_2}{X_0}\frac{\sinh{\paran{\beta X_0}}}{\cosh{\paran{\beta X_0}}+
\exp{\paran{-2\beta J_z}}\cosh{\paran{\beta Y_0}}},} where
\eq{denk8}{
X_0=\left[\paran{J_x-J_y}^2+(x+y+H_1+H_2)^2\right]^{1/2}, \quad Y_0=\left[\paran{J_x+J_y}^2+(x-y+H_1-H_2)^2\right]^{1/2}
.}

The coefficients in Eq. \re{denk6} are defined by
\eq{denk9}{
\begin{array}{lcl}
A_{x}=\cosh{\paran{J_z\nabla_x}}&\quad&
B_{x}=\sinh{\paran{J_z\nabla_x}}\\
A_{y}=\cosh{\paran{J_z\nabla_y}}
&\quad&
B_{y}=\sinh{\paran{J_z\nabla_y}}\\
A_{xy}=\cosh{\left[J_z\paran{\nabla_x+\nabla_y}\right]}&\quad&
B_{xy}=\sinh{\left[J_z\paran{\nabla_x+\nabla_y}\right]},\\
\end{array}
}
where $\nabla_x=\partial/\partial x$ and $\nabla_y=\partial/\partial y$ are
the usual differential operators in the differential operator technique.
Differential operators act on an arbitrary function $g$
via \eq{denk10}{\exp{\paran{a\nabla_x+b\nabla_y}}g\paran{x,y}=g\paran{x+a,y+b}} for
arbitrary constants  $a$ and $b$.

With using Eq \re{denk9} in  \re{denk6} and performing Binomial
expansion we can obtain the expression as polynomial in $m$.
If we place this resulting expression in Eq. \re{denk3} we can
obtain dynamical equation of motion as \eq{denk11}{\theta \tur{m}{t}=-m+\summ{k=0}{z}{}C_{k}m^{k},}
where the coefficients are defined by \eq{denk12}{
C_{k}=\summ{p=0}{z_0}{}\summ{q=0}{z_0}{}\summ{r=0}{z_1}{}\delta_{p+q+r,k}C^\prime_{pqr}}
where $\delta_{i,j}$ is the Kronecker delta and \eq{denk13} {C^\prime_{pqr}=\komb{z_0}{p}\komb{z_0}{q}\komb{z_1}
{r}A_x^{z_0-p}A_y^{z_0-q}A_{xy}^{z_1-r}B_x^{p}B_y^{q}B_{xy}^{r}f\paran{x,y,H_1,H_2}|_{x=0,y=0}.}

If we assume that the system is under the influence of an spatially homogenous magnetic
field (i.e. $H_1(t)=H_2(t)=H(t)$), for a given set of Hamiltonian
parameters ($J_x,J_y,J_z,H_0,\omega$), as well as temperature,  by determining
the coefficients  from Eq. \re{denk12} we can obtain equation of motion
Eq. \re{denk11}. This differential equation can be solved numerically.
We use RK4 (fourth order Runge-Kutta  method) for the solution of the Eq. \re{denk11}.
This iterative method starts some initial value of the magnetization ($m(0)$) and arrive the
wanted solution after the convergency criteria $m\paran{t}=m\paran{t+2\pi/\omega}$ satisfied.
By this way we can obtain DOP as

\eq{denk14}{Q=\frac{\omega}{2\pi}\intego{}{}{m\paran{t}dt}} where $m\paran{t}$
is a stable and periodic function. Two kind of solutions occur in these systems:
Symmetric solutions which corresponds to the paramagnetic phase (P) and
nonsymmetric solutions which corresponds to the ferromagnetic phase (F).
Symmetric solutions satisfy the property \eq{denk15}{m\paran{t}=-m\paran{t+\pi\omega}}
while  nonsymmetric solutions do not satisfy of Eq. \re{denk15}.
Although P and F solutions of Eq. \re{denk11} can be obtained by
any choice of the initial value $m(0)$, there can be some situations
which the stable solutions depends on the choice of the initial
value $m(0)$. Phase related to this solutions called mixed phase (F+P).
Dynamical critical points can be determined by obtaining the variation of the $Q$
with temperature for given set of Hamiltonian parameters.

We can construct the hysteresis loops which are nothing but the variation
of the $m(t)$ with $H(t)$ in one period of the periodic magnetic field.
Hereafter,  once the hysteresis loop is determined, some quantities about
it can be calculated. One of them is  dynamical HLA and can be calculated
via integration over of the one period of the magnetic field,
\eq{denk16}{A=\intego{}{}{}m\paran{t}dH} and corresponds to the energy
loss due to the hysteresis. Another two quantity which can describe the
shape of the hysteresis loop can be determined, namely CF and RM.

\section{Results and Discussion}\label{results}

In order to focus on the effect of the anisotropy in the exchange interaction
on the DPT and hysteresis characteristics of the system, let us choose $J_x=J_y$ and scale
this quantity as well as the temperature and amplitude of the magnetic field   with the unit of energy $J_z$ as,
\eq{denk17}{\Delta=\frac{J_x}{J_z}=\frac{J_y}{J_z}, \tau=\frac{k_BT}{J_z}, h_0=\frac{H_0}{J_z},h(t)=\frac{H(t)}{J_z}.
} Our investigation will be for simple cubic  ($z_0=5,z_1=0$) lattice. From Eq. \re{denk17}
we can say that, $\Delta=0$ corresponds to the Ising model, while $\Delta$ rises starting from zero,
system arrives to the isotropic Heisenberg model ($\Delta=1$) with
passing through the  XXZ type symmetric  Heisenberg model ($0<\Delta<1$).

The physical mechanisms giving rise to the DPT, as well as the influences of the
amplitude and frequency of the magnetic field on the hysteresis behaviors are well known.
Detailed explanations can be found in Refs. \cite{ref1,ref32}.
So,  we want to focus only on the effect of the anisotropy in the exchange
interaction on both phase diagrams and hysteresis behaviors.
We set the value of $\theta=1$ throughout our numerical calculations.
\newpage

\subsection{Dynamic Phase Boundaries}

Variations of the dynamical critical temperatures ($\tau_c$) with the amplitude
of the magnetic field ($h_0$) can be seen in Fig. \re{sek1},  for isotropic Heisenberg model (curves labeled by B)
in comparison with the Ising model (curves labeled by A). Dynamic phase boundaries (DPBs) separating
dynamically ordered and disordered phases have been drawn for  two selected
values of applied field frequency $\omega=0.5$ in Fig. \re{sek1} (a)
and  $\omega=5.0$ in Fig. \re{sek1} (b), respectively. One can clearly see from the
Fig. \re{sek1} that DPB related to the Ising case lies above of the
isotropic  Heisenberg case which means that rising anisotropy
gives rise to shift the dynamical critical point to upwards
in the  $(h_0-\tau_c)$ plane. For a fixed sets of amplitude and frequency of
the applied field, when the anisotropy increases, the spins tend to
align in $z$ direction, hence, more thermal energy is needed to broke
the aforementioned spin alignment and to observe a DPT. In accordance with the expectations,
for value of $h_0=0.0$ corresponding to the static Ising model for the simple cubic
lattice, both of the DPBs regarding to the Ising model start with the same critical
temperature $\tau_c= 5.039$. The same situation is also valid for the phase diagrams of the isotropic
Heisenberg model, at zero magnetic field amplitude $\tau_c=4.891$. An
increment in value of $h_0$ causes to decline the critical temperature. In the regions of
higher values of the amplitude and lower values of the temperature, dynamic first order
transitions appear. Furthermore, the region which is limited by this first order line (dotted lines in Fig. \re{sek1})
and second order line, F+P phases appear, as explained in detail in  the Sec. \re{formulation}.
Time dependent magnetization of the system strongly depends on the selected value of the initial
magnetization ($m(0)$) in coexistence region. In order to show these types
of treatments, in Fig. \re{sek2} we choose two representative behaviors of the magnetizations
with time (t), for the isotropic Heisenberg model. The frequency is chosen as $\omega=5.0$ and
the other parameters are chosen as such that give rise to existence of a F+P phase (Fig. \re{sek2} (a))
and  F phase (Fig. \re{sek2} (b)). For each plot, two initial values of magnetization,
namely $m(0)=0.2$ (curve labeled by A) and $m(0)=0.8$ (curve labeled by B) are selected.
It can be easily seen from the Fig. \re{sek2} (a) that depending upon the selected values of
initial condition, the time dependent magnetizations oscillate around different magnetization values
during the RK4 iteration indicating a F+P phase, whereas this is not the case for Fig. \re{sek2} (b).

Moreover,  as seen in Fig. \re{sek1} that, the considered magnetic system exhibits
a dynamic tricritcal point (DTCP) where dynamic first and second order phase transition
lines meet. One can conclude from Fig. \re{sek1} that,  when the value of the applied field
frequency increases, the temperature coordinates of the DTCPs shift to upper
region for both Ising and Heisenberg models. It should be noted here that the
aforementioned situation is consistent with the results obtained in Ref. \cite{ref30}
where dynamic phase transition properties of the classic Heisenberg model is
analyzed by making use of MFA. It is also beneficial to adress that temperature
coordinate of the DTCP of Ising model seems to be above
point of the isotropic Heisenberg model for considered values of frequencies.

By the way, we want to underline an important point concerning of DPBs.
For a considered value of $\Delta$, we have three parameters affecting the dynamic critical
nature of the system, and  if one keeps the two of them fixed, inducing DPT
comes from the third one. So, based on Fig. \re{sek1} it can be said that critical
temperature as well as critical amplitude values of the Ising model are
higher than the isotropic Heisenberg model.
We can say  following the same analogy that, the value of the critical frequency of the Ising model
is lower than the isotropic Heisenberg model. This means that,
rising anisotropy in the exchange interaction causes to
decline of the critical frequency value.

\begin{figure}[h]\begin{center}
\epsfig{file=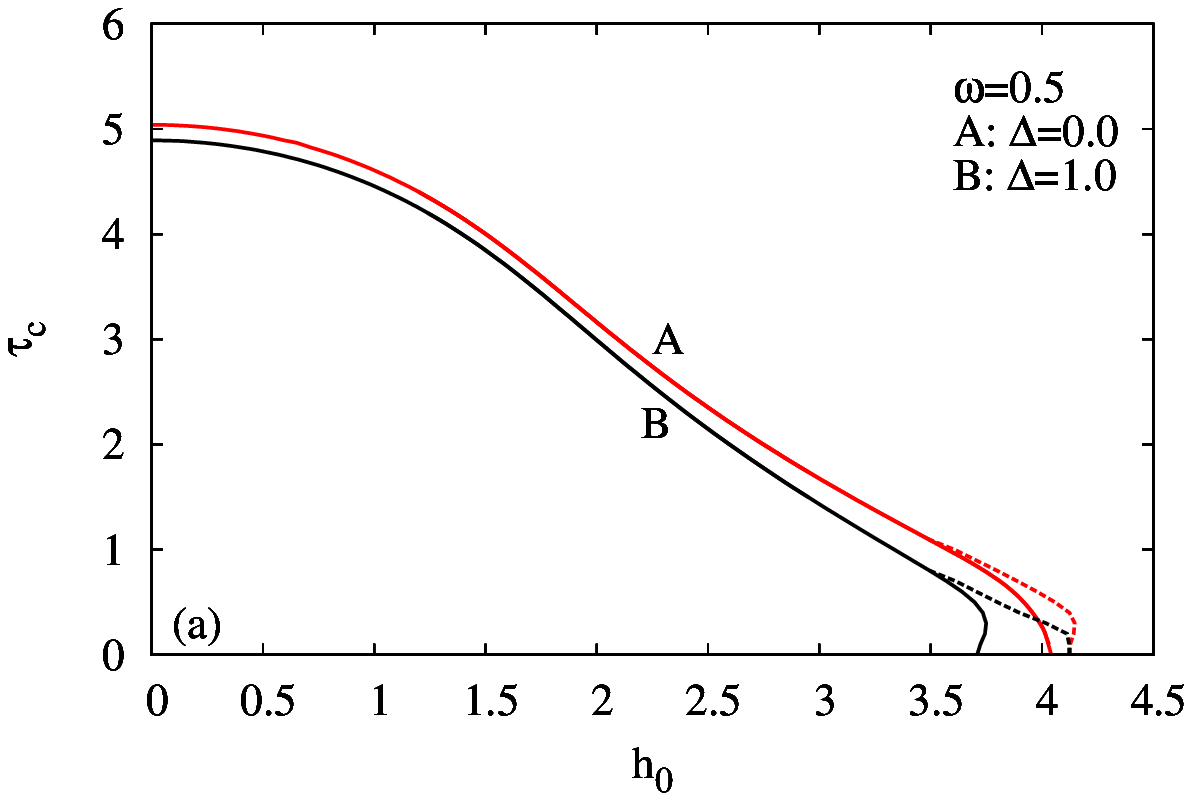, width=7.2cm}
\epsfig{file=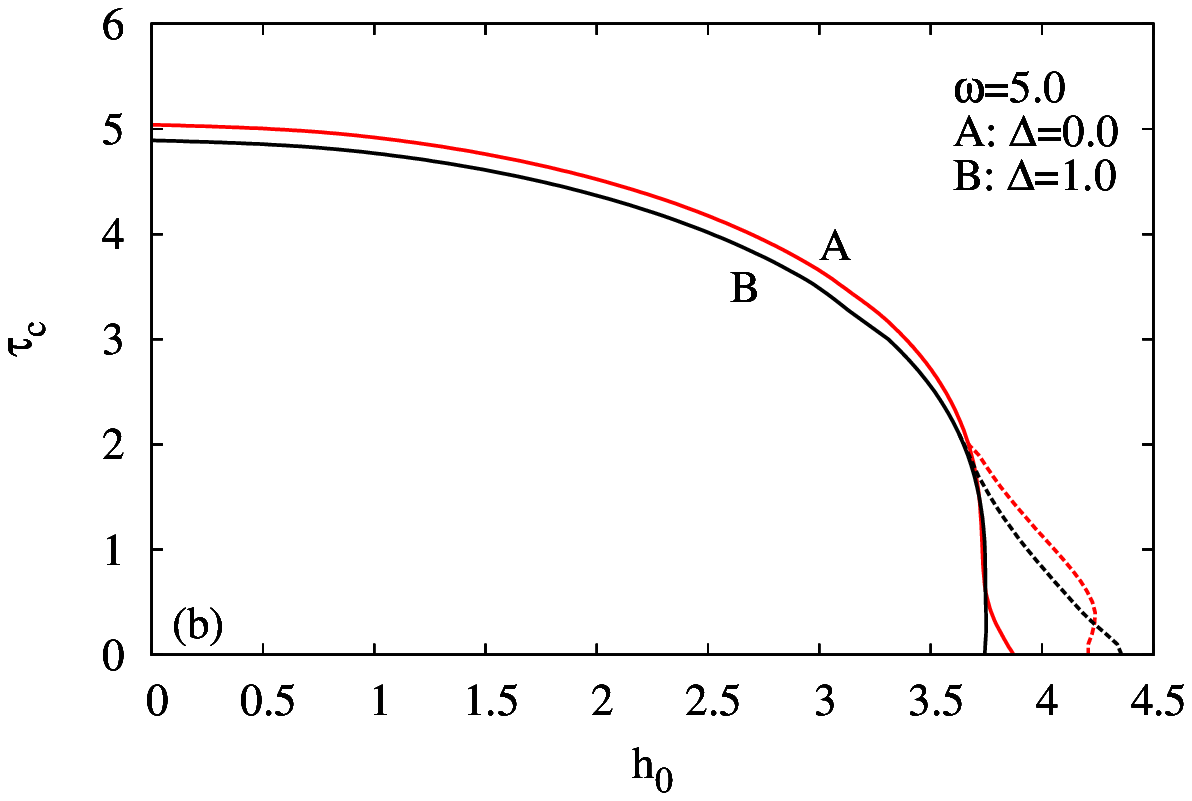, width=7.2cm}
\end{center}
\caption{Variation of the critical temperature with the amplitude of the
magnetic field for the Ising model (curves labeled with A, red curve) and
isotropic Heisenberg model (curves labeled with B, black curve)  for selected
values of frequency (a) $\omega=0.5$ and (b) $\omega=5.0$. Solid line
represents the second order transitions, while the dotted one corresponds
to the first order transitions.} \label{sek1}\end{figure}

\begin{figure}[h]\begin{center}
\epsfig{file=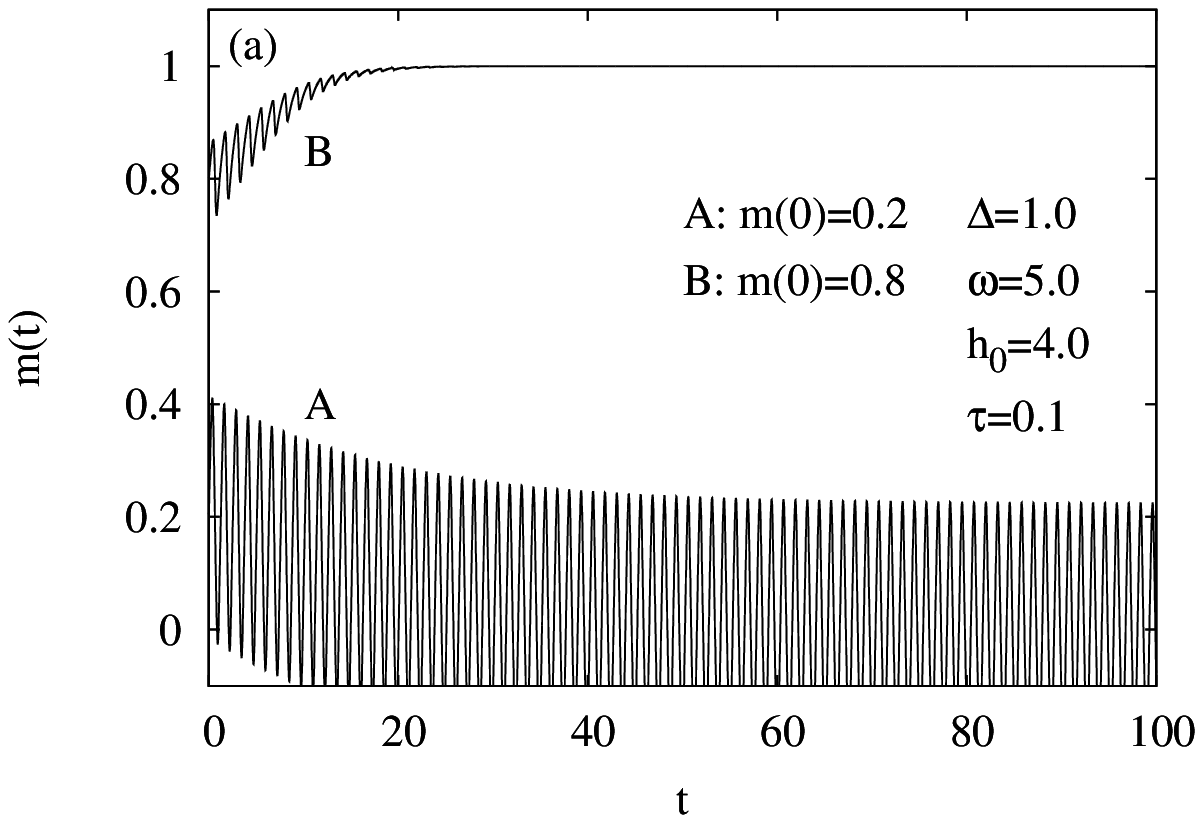, width=7.2cm}
\epsfig{file=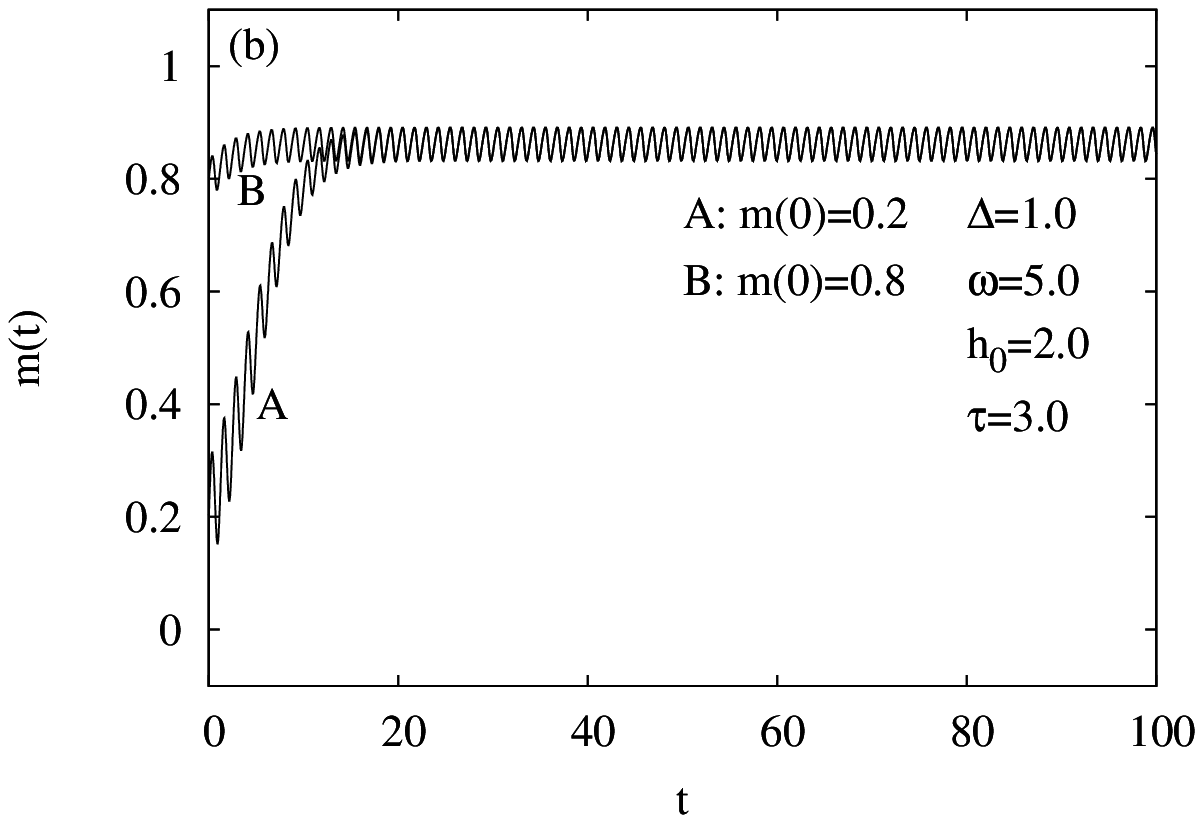, width=7.2cm}
\end{center}
\caption{Variation of the magnetization with time for parameters that
belong (a) F+P phase and (b) F phase. Considered values of parameters as well as
initial values of magnetizations are as shown in the plots.} \label{sek2}\end{figure}

\subsection{Hysteresis Caharacteristics}

In Fig. \re{sek3} we give  the frequency variation ($\omega$) of the HLA for the parameters
mentioned above.  All curves except from the curve labeled by A in
Fig. \re{sek3} (a) are related to the dynamically paramagnetic phase.
For the values of $\tau=2.0$ and $h_0=3.0$ the system passes
to the ferromagnetic phase at $\omega_c=0.765$ which shows itself in the
curve labeled A in Fig. \re{sek3} (a). For the dynamically paramagnetic phase, 
when $\omega$ increases,  value of HLA increases starting from a certain value depending on 
the system parameters and shows a frequency induced maximum, then, it begins to fall. 
For the curve labeled by A in Fig. \re{sek3} (a), HLA falls faster than the others
for the values of frequency that provide $\omega>\omega_c$, due to 
the occurence of the paramagnetic-ferromagnetic DPT. Besides, the
curves related to the higher amplitudes are above of others (e.g. 
compare curves labeled by C and B in Fig. \re{sek3} (b)). The other
typical situation is, maximum value of HLA of the system with higher
temperature is lower than the system with lower temperature (e.g. 
compare curves labeled by C in Figs. \re{sek3} (a) and (b)).   	

\begin{figure}[h]\begin{center}
\epsfig{file=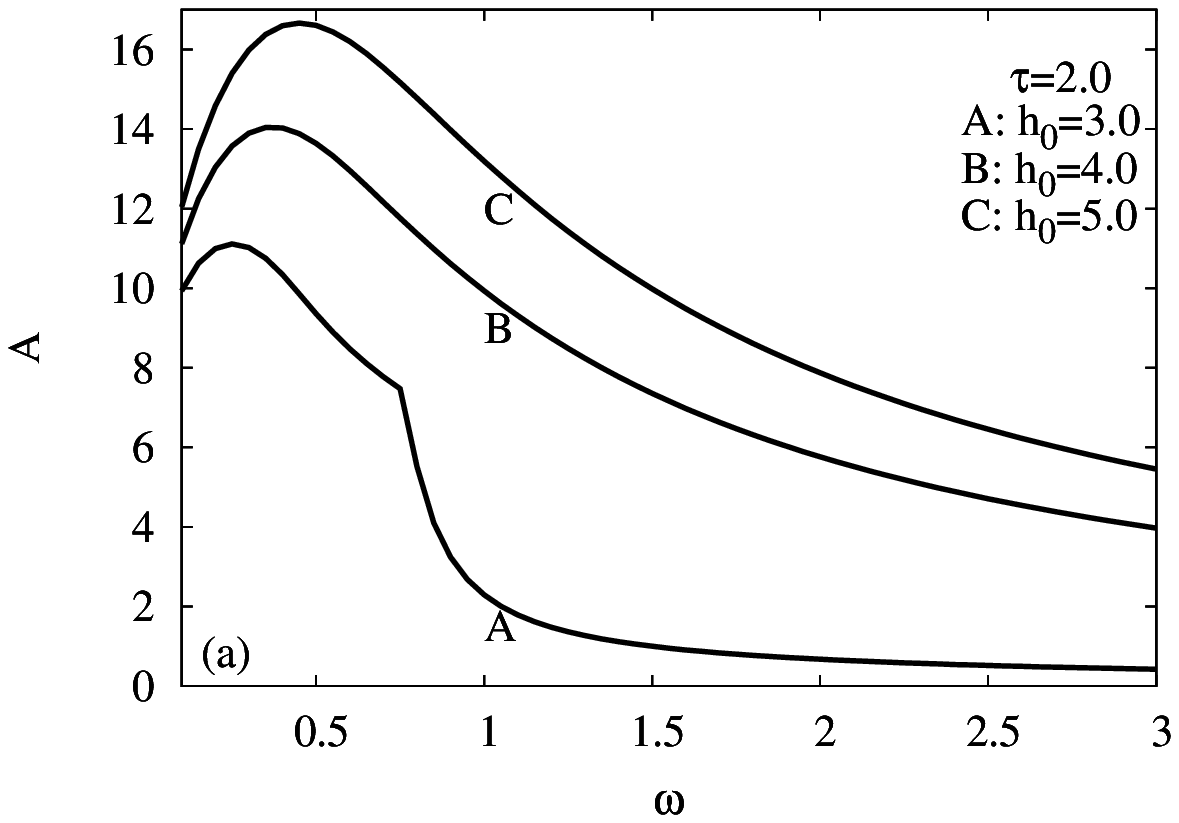, width=7.0cm}
\epsfig{file=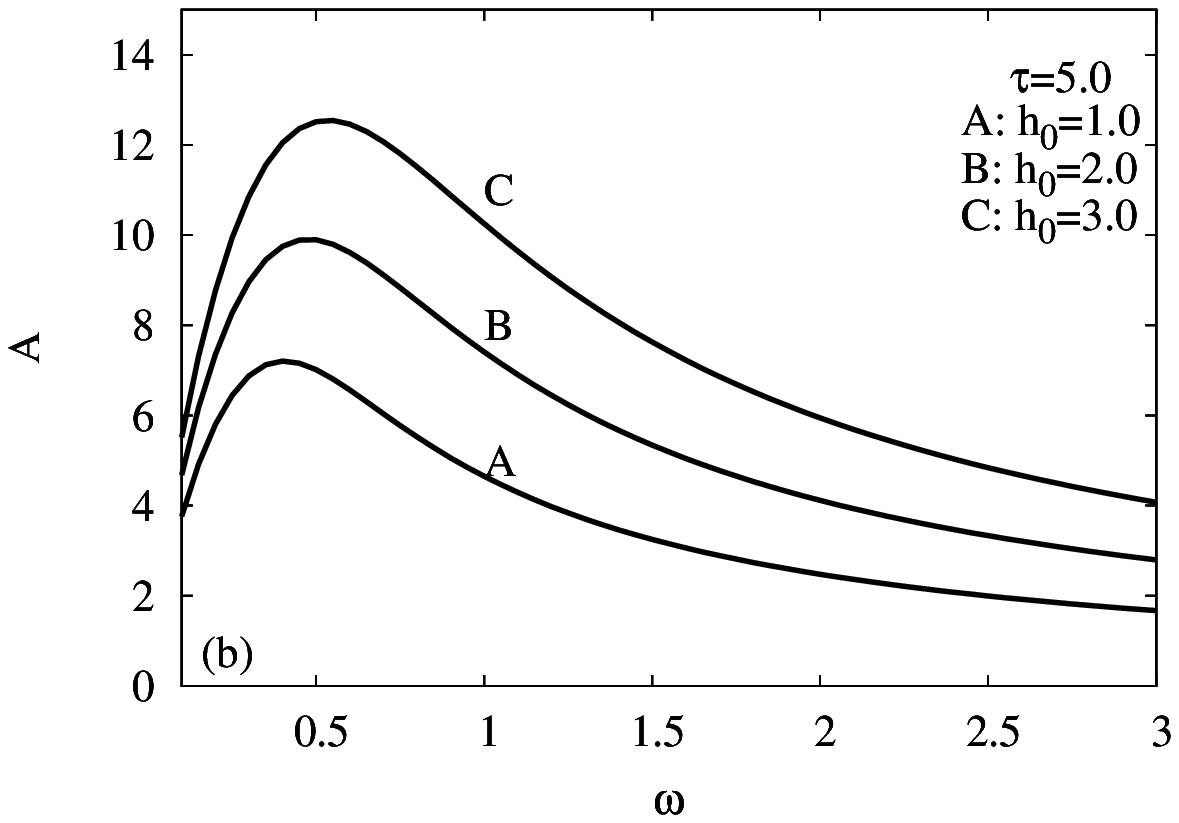, width=7.0cm}
\end{center}
\caption{Variation of the HLA with frequency of the magnetic field, 
for selected values of magnetic field amplitudes and  temperatures 
(a) $\tau=2.0$ and  (b) $\tau=5.0$. Selected value of $\Delta=1.0$, i.e. 
isotropic Heisenberg model.} \label{sek3}\end{figure}

Frequency variation of  CF, for the  parameter values mentioned above, can be seen in Fig. \re{sek4}. 
Again, occurrence of the DPT at a value of $\omega_c=0.765$ shows itself in the related curve, i.e. 
curve labeled by A in Fig. \re{sek4} (a). Typical behavior of the  CF with rising frequency is that 
it first rises for a while after than getting constant at a value of $h_0$, except the curve labeled 
by A in Fig. \re{sek4} (a).  As seen in Figs. \re{sek4} (a) and (b) this behavior is valid 
both of the chosen temperatures. For the curve labeled by A in Fig. \re{sek4} (a), some after 
the value of $\omega_c$, CF gets the value of zero.   For the values of $\omega>\omega_c$ system 
is in the ferromagnetic phase, i.e. DOP is different from zero. But this does not mean that, 
the time dependent magnetization can not have the negative values in one period of the 
magnetic field. In other words $m(t)<0$ can be satisfied while $Q>0$ according to the 
Eq.  \re{denk14}. If we look at the curve labeled by A in  Fig. \re{sek4} (a) carefully,
we can see this situation. At the value of $\omega=\omega_c$ CF is not zero, it goes to 
zero between $\omega_c=0.765$ and $\omega=0.840$ when $\omega$ rises, finally after 
the value of  $\omega=0.840$, rising frequency cannot change the zero value of CF.

\begin{figure}[h]\begin{center}
\epsfig{file=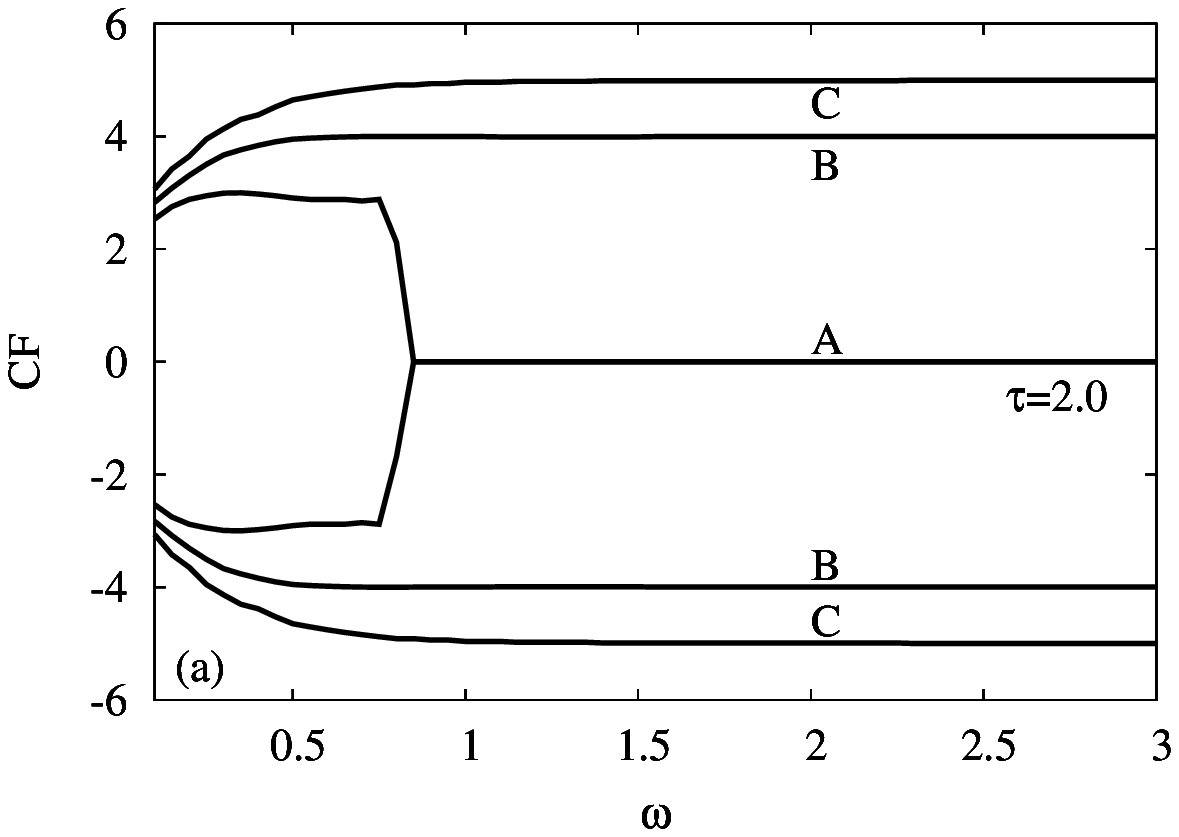, width=7.0cm}
\epsfig{file=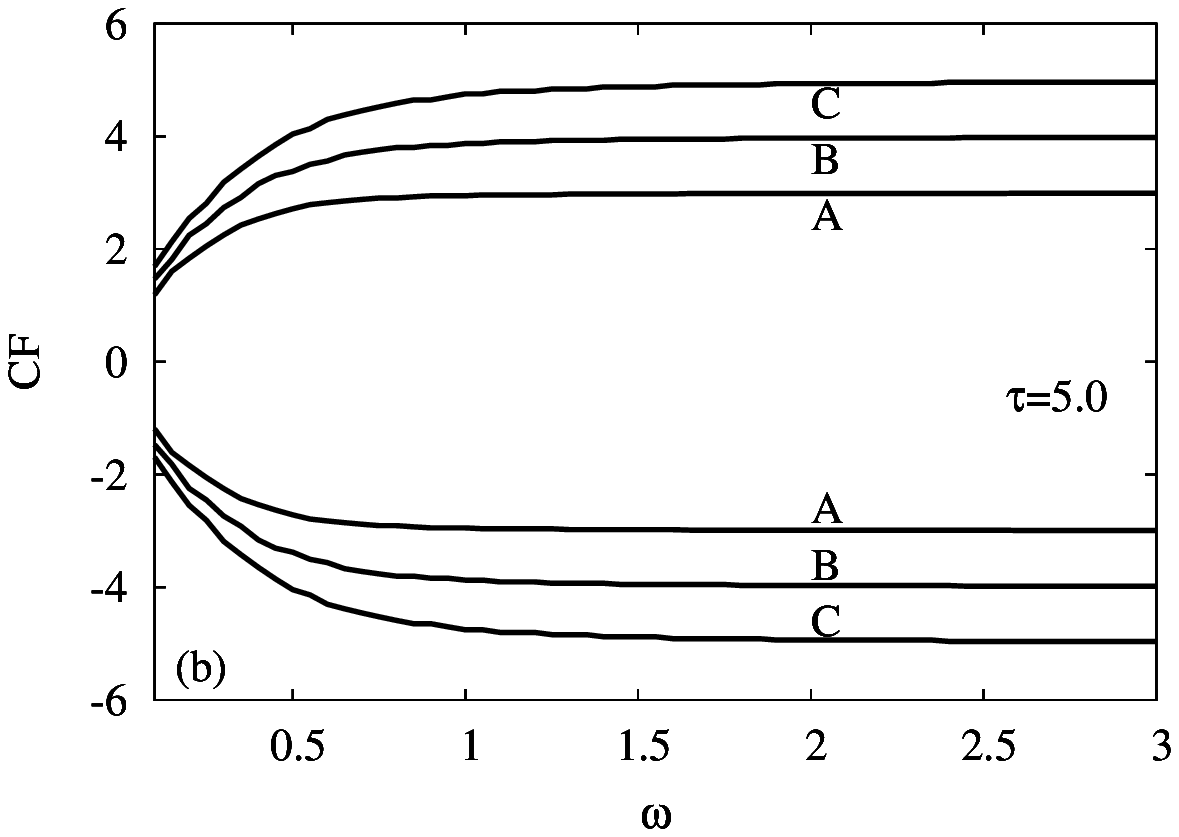, width=7.0cm}
\end{center}
\caption{Variation of the CF with frequency of the magnetic field, for selected 
values of magnetic field amplitudes and  temperatures (a) $\tau=2.0$ and  (b) $\tau=5.0$. 
Selected value of $\Delta=1.0$, i.e. isotropic Heisenberg model. Curve labels (A,B,C) are 
related to same values mentioned in Fig. \re{sek3}.} \label{sek4}\end{figure}

Variation of RM with frequency, for the values of parameters mentioned above can be seen in Fig. \re{sek5}. 
In general we can talk about two different behavior for the paramagnetic RM curves. For lower temperatures 
(Fig. \re{sek5} (a), except the curve labeled by A),  rising $\omega$ cannot change the value of RM, 
which is $\pm 1.0$ for a while. After than, rising frequency decreases the RM. In contrast to 
this, for higher temperatures (Fig. \re{sek5} (b)), rising frequency first slightly rises RM, 
after it reaches the maximum value, starts to decrease while $\omega$ rises. The curve 
labeled by A in Fig. \re{sek5} (a) is related to the situation where the DPT occurs from the 
paramagnetic phase to the ferromagnetic one, with rising frequency of the magnetic field. 
The trend mentioned above changes at a value of $\omega_c$ as seen in curve labeled by A Fig. \re{sek5} (a). 
Two branch of the RM curve shift upwards, means that hysteresis loops shifts upwards in 
the $(m(t)-h(t))$ plane for the rising frequency that provide $\omega>\omega_c$.

\begin{figure}[h]\begin{center}
\epsfig{file=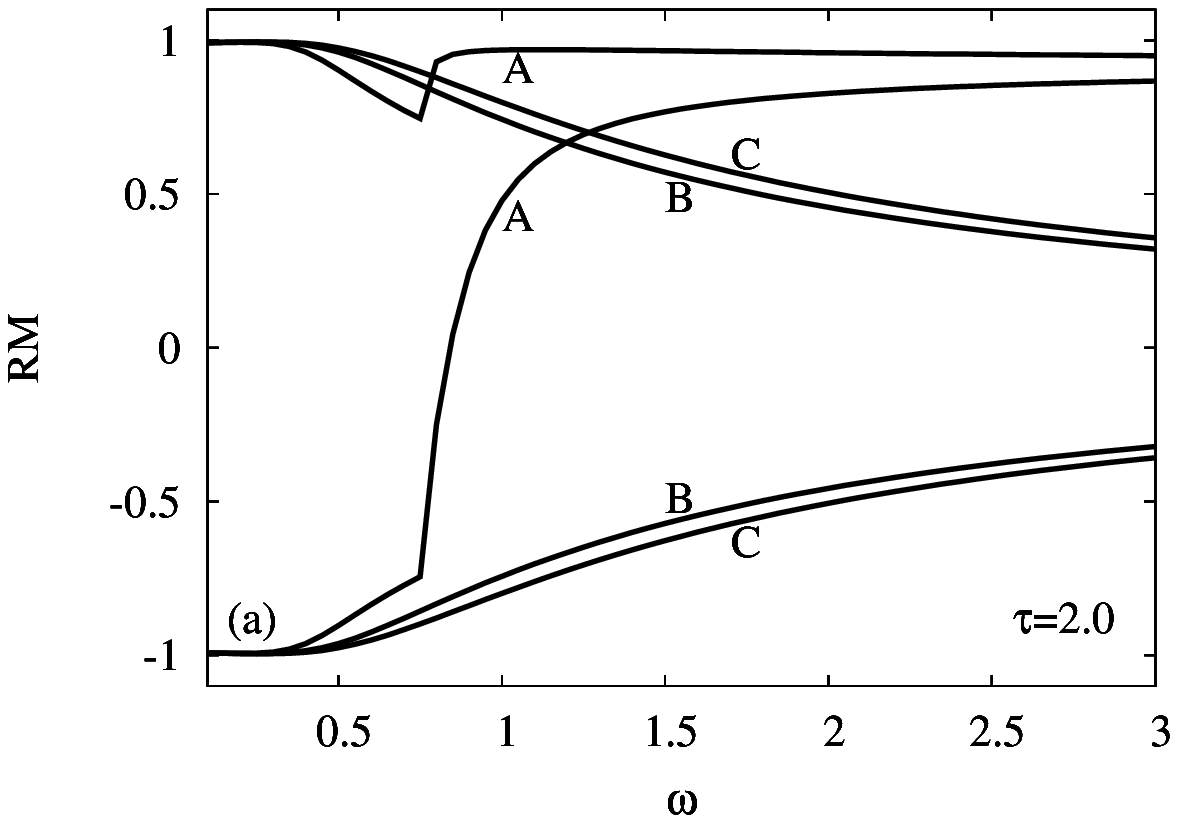, width=7.0cm}
\epsfig{file=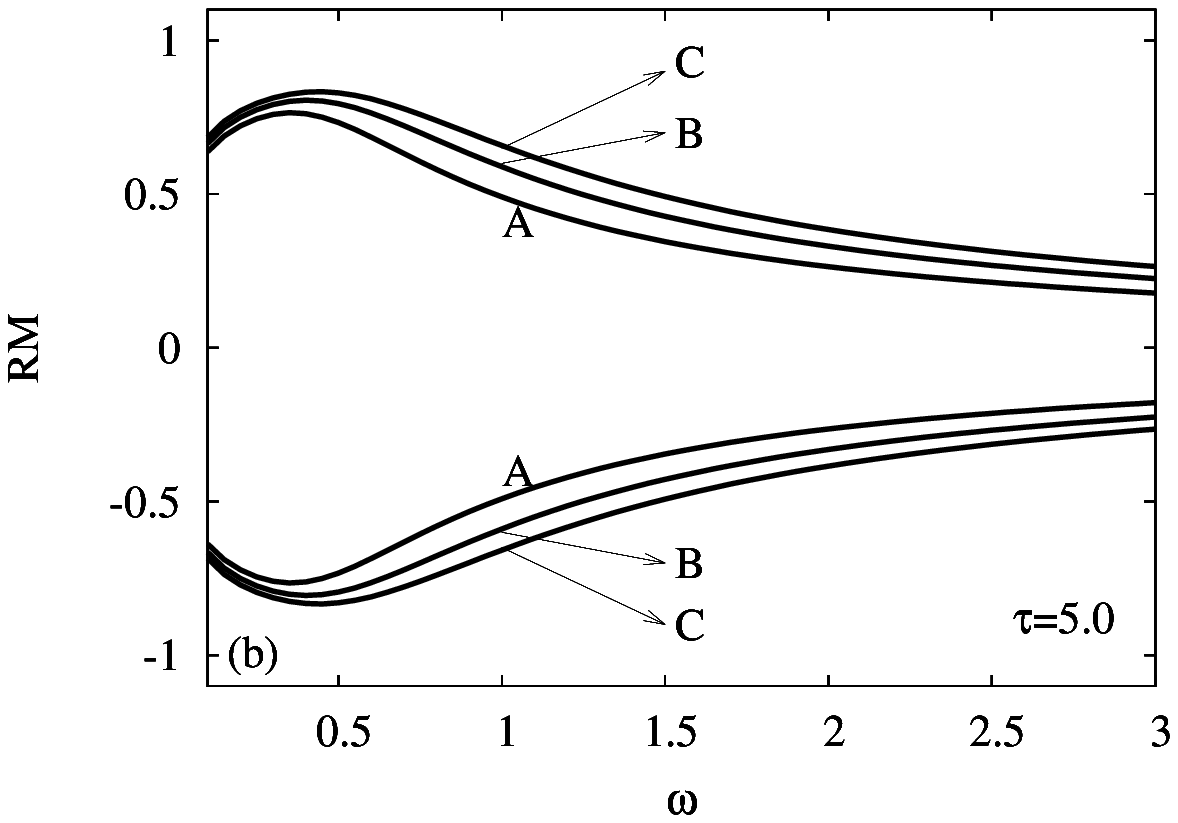, width=7.0cm}
\end{center}
\caption{Variation of the RM with frequency of the magnetic field, for selected 
values of magnetic field amplitudes and  temperatures (a) $\tau=2.0$ and  (b) $\tau=5.0$. 
Selected value of $\Delta=1.0$, i.e. isotropic Heisenberg model. Curve labels (A,B,C) are 
related to same values mentioned in Fig. \re{sek3}.} \label{sek5}\end{figure}

Behavior of the curves labeled by A in Figs. \re{sek3} (a), \re{sek4} (a) and \re{sek5} (a),  
can be seen more clearly from Fig. \re{sek6}, which is hysteresis loops 
for the $\tau=2.0, h_0=3.0$ and selected values of frequency around the 
value of $\omega_c$. The curves related to the frequency values $\omega=0.74,0.76$ 
are symmetric about the origin and they are related to the paramagnetic phase. 
All other curves are related to the ferromagnetic phase. As seen in Fig. \re{sek6}, 
rising frequency results the hysteresis loops shift upwards. Altough the system is 
in the ferromagnetic phase, CF is nonzero until the frequency reaches to the 
value of $0.84$. After this value CF is zero. Similar situation is valid for 
the RM. For the values of $\omega<0.84$ RM has two values which are one of 
is positive and the other one is negative. For the loop related to 
the $\omega=0.84$, lower value of the RM gets the value of zero and 
after this value, rising frequency results both of the RM values positive.

\begin{figure}[h]\begin{center}
\epsfig{file=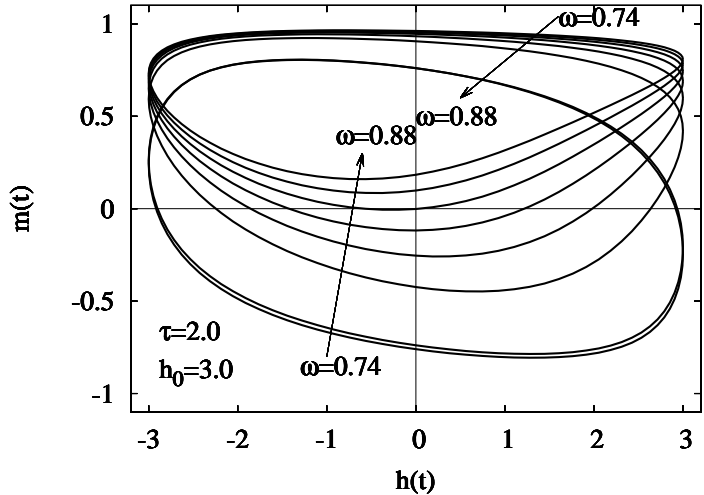, width=10.0cm}
\end{center}
\caption{Hysteresis loops for the isotropic Heisenberg model ($\Delta=1.0$) for 
selected values of frequency of the magnetic field and  (a) $\tau=2.0$, $h_0=3.0$. 
The frequency values start from the $0.74$ and end with $0.88$ in the direction 
shown by arrows, with increments $0.2$. } \label{sek6}\end{figure}

After this short conclusions about the hysteresis loop behaviors with 
rising frequency of the magnetic field, let us look at the effect 
of the rising anisotropy in the exchange interaction on the behavior 
of the hysteresis loops. The variation of the HLA with frequency can 
be seen in Fig. \re{sek7} for the values of $\tau=2.0, h_0=3.0$ and $\Delta=0.0,0.5,1.0$ i.e.   
Ising model, XXZ model (with $\Delta=0.5$) and isotropic Heisenberg model, respectively.  
As seen in the Fig. \re{sek7}, the behavior of the HLA with the rising frequency 
is similar for the three of the model. All of the models show phase transitions 
at a specific values of the frequency, namely $\omega_c=0.670$ for the Ising 
model, $\omega_c=0.695$ for the XXZ model (with $\Delta=0.5$) and $\omega_c=0.765$ for the isotropic Heisenberg model. Beside the similarity of the behaviors, one important situation happened. As seen  Fig. \re{sek7} that, for lower 
frequency values, when the anisotropy in the exchange interaction rises
HLA increases, while for the higher values of the frequency, this situation 
getting reverse. In other words, the value of HLA of the Ising model is 
higher than that of the isotropic Heisenberg model, while after the
value of $\omega^*=0.312$  HLA of the isotropic Heisenberg model is 
higher than that of the Ising  model. We will talk more 
about this situation below.

\begin{figure}[h]\begin{center}
\epsfig{file=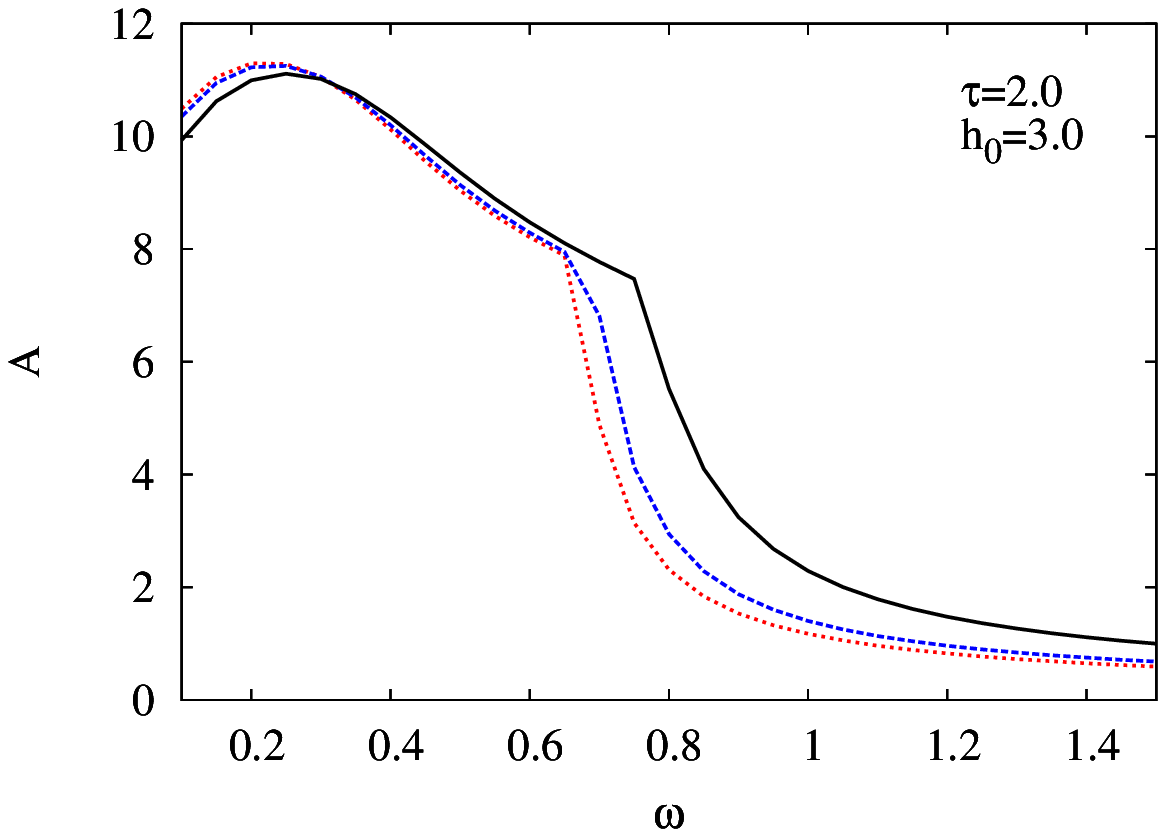, width=7.0cm}
\end{center}
\caption{Variation of the HLA with frequency of the magnetic field, for the 
values of $\tau=2.0$ and $h_0=3.0$. The dotted line (red curve) corresponds 
to the Ising model ($\Delta=0.0$), dashed line (blue curve) corresponds to 
the XXZ type anisotropic Heisenberg model (with $\Delta=0.5$), while solid 
line (black curve) corresponds to the isotropic Heisenberg 
model ($\Delta=1.0$).} \label{sek7}\end{figure}

The same situation is valid for the behavior of the CF. The variation of 
the CF with the frequency can be seen in Fig. \re{sek8} (a) for three of 
the model. For lower values of the frequency CF of the Ising model is 
slightly higher than that of the isotropic Heisenberg model. After the 
value of $\omega^*=0.312$, CF of the isotropic Heisenberg model getting 
higher than that of the Ising model. CF of the XXZ model always lies 
between these two CF values, namely CF of the Ising model and isotropic 
Heisenberg model. The difference of the critical frequency values of the 
models that have different anisotropies shows itself in the values of the 
frequencies, where CF starts to fall zero. On the other hand, 
in Fig. \re{sek8} (b) we can see the variation of the RM with 
frequency for these three models. Two branch of the RM for 
different models are almost the same for the low frequency 
region. After the DPT frequency of the Ising model, RM of the 
Ising model lies above of the two others, while for the higher 
frequencies these three curves approach each other.

\begin{figure}[h]\begin{center}
\epsfig{file=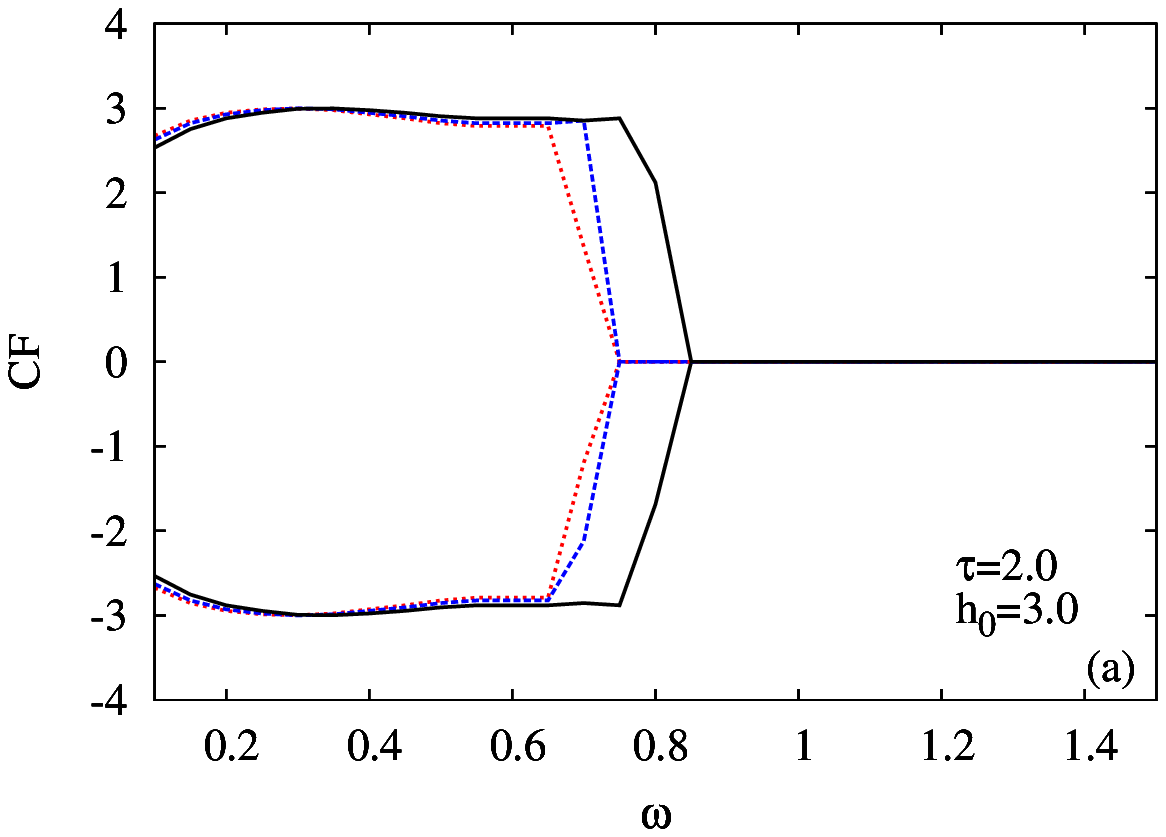, width=7.0cm}
\epsfig{file=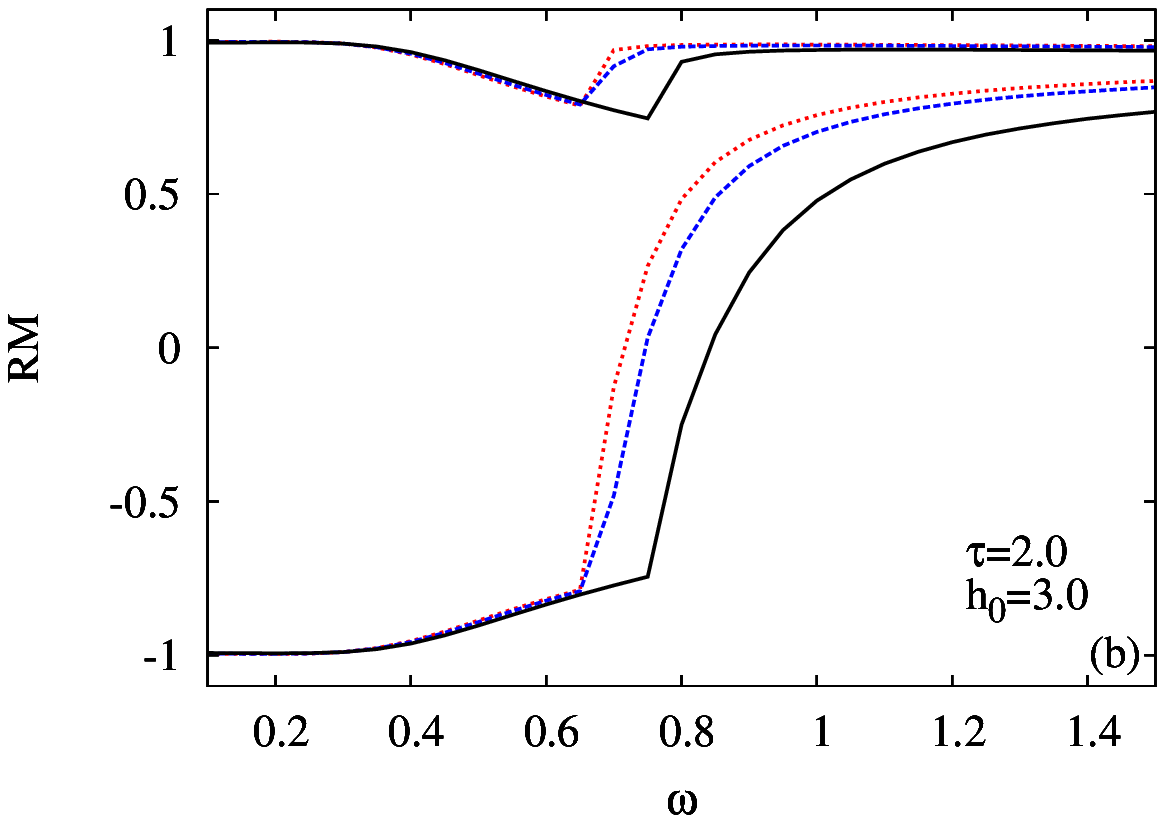, width=7.0cm}
\end{center}
\caption{Variation of the (a) CF and (b) RM  with frequency of the magnetic 
field, for the values of $\tau=2.0$ and $h_0=3.0$. The dotted line (red curve) 
corresponds to the Ising model ($\Delta=0.0$), dashed line (blue curve) 
corresponds to the XXZ type anisotropic Heisenberg model (with $\Delta=0.5$),
while solid line (black curve) corresponds to the isotropic 
Heisenberg model ($\Delta=1.0$). } \label{sek8}\end{figure}

All these observations about the effect of the anisotropy in the exchange 
interaction can be seen more clearly in the hysteresis loops. In Fig. \re{sek9} 
we depict the hysteresis loops of the three aforementioned models, 
for $\tau=2.0$ and $h_0=3.0$, with selected frequency values. For 
the value of $\omega=0.2$, Ising hysteresis loop lies outside of  
the others (Fig. \re{sek9} (a)), means that HLA of the Ising hysteresis 
loop is the greatest one. When the frequency rises, the shape of the 
loops change and Ising loop starts to settle inside to the other 
hysteresis loops (Fig. \re{sek9} (b) and (c)). Up to now all models 
are in the paramagnetic phase. While the frequency rises, after the 
value of $\omega=0.670$ Ising model passes to the ferromagnetic phase, 
the hysteresis loop for the Ising model starts to move 
upward (Fig. \re{sek9} (d)). Similarly  after the value 
of $\omega=0.695$, XXZ model (with $\Delta=0.5$) passes to 
the ferromagnetic phase (Fig. \re{sek9} (e)). Finally, 
since the  $\omega=0.8>0.765$ (which is the critical 
frequency of the isotropic Heisenberg model)  
all loops in Fig. \re{sek9} (f) are related to the ferromagnetic phase.
	
\begin{figure}[h]\begin{center}
\epsfig{file=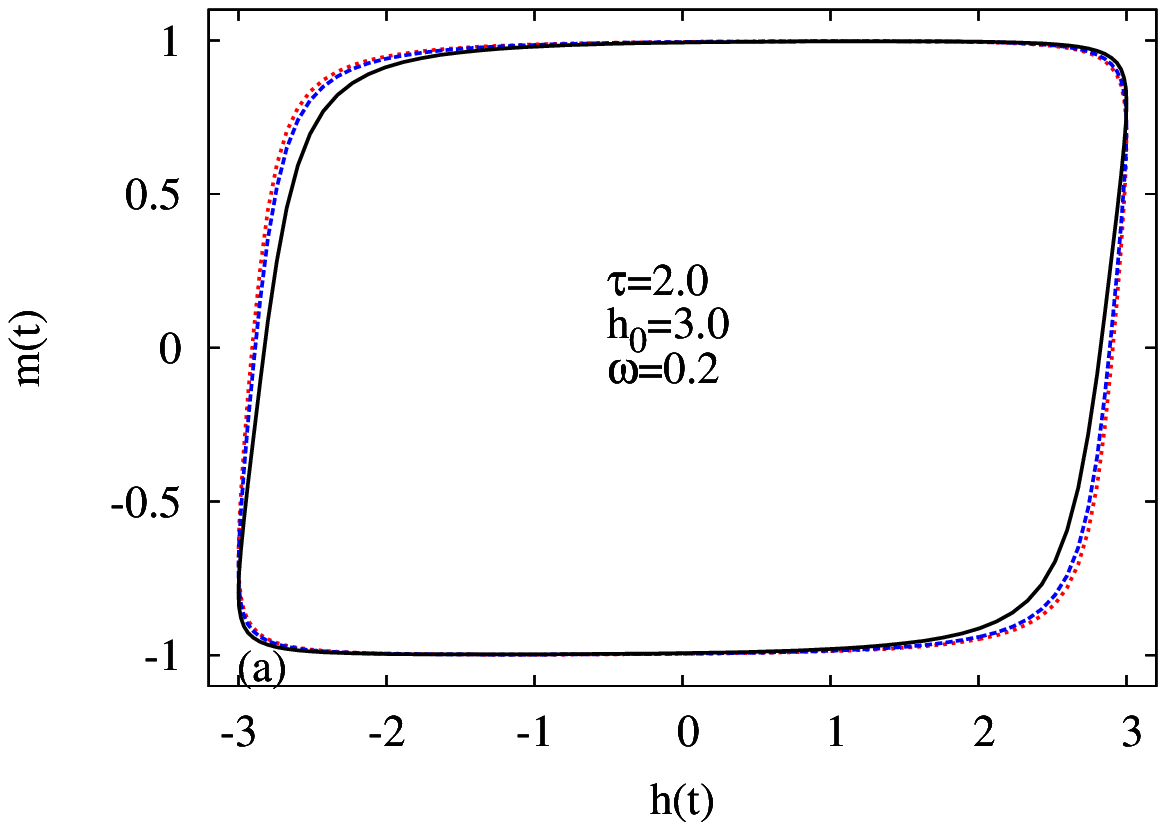, width=7.0cm}
\epsfig{file=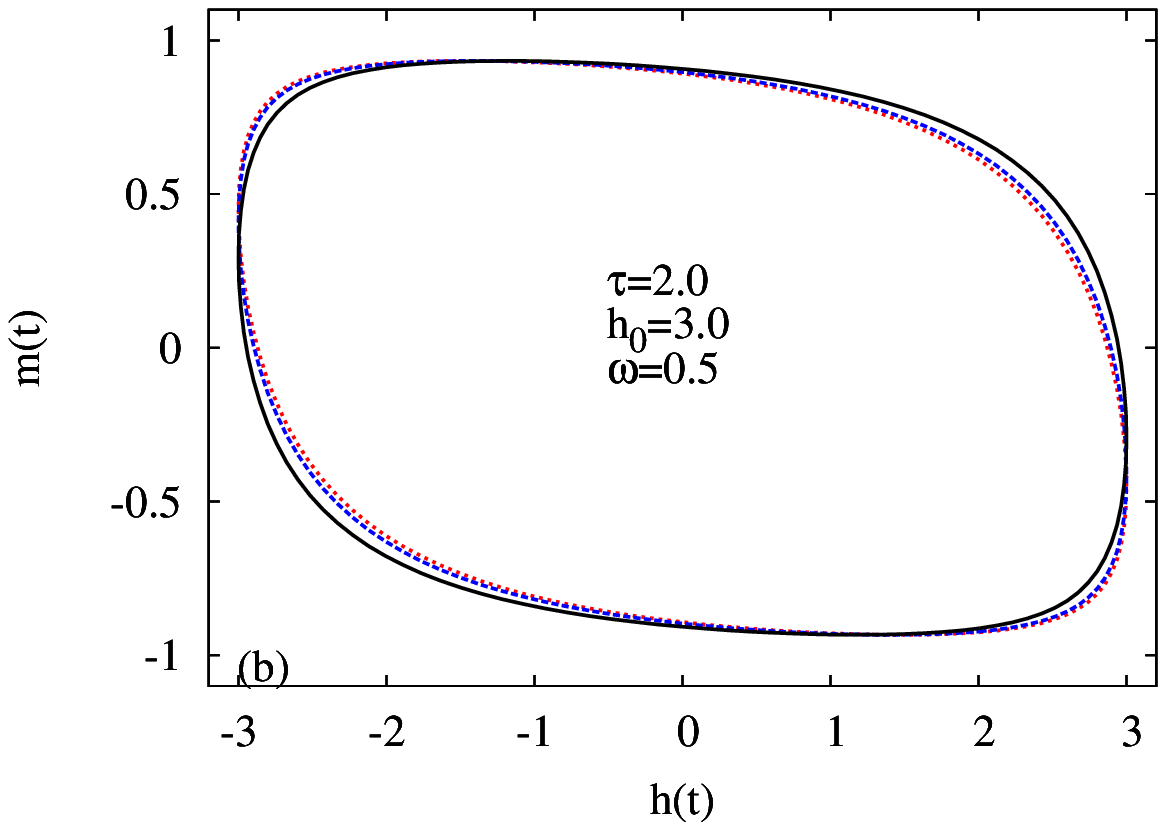, width=7.0cm}
\epsfig{file=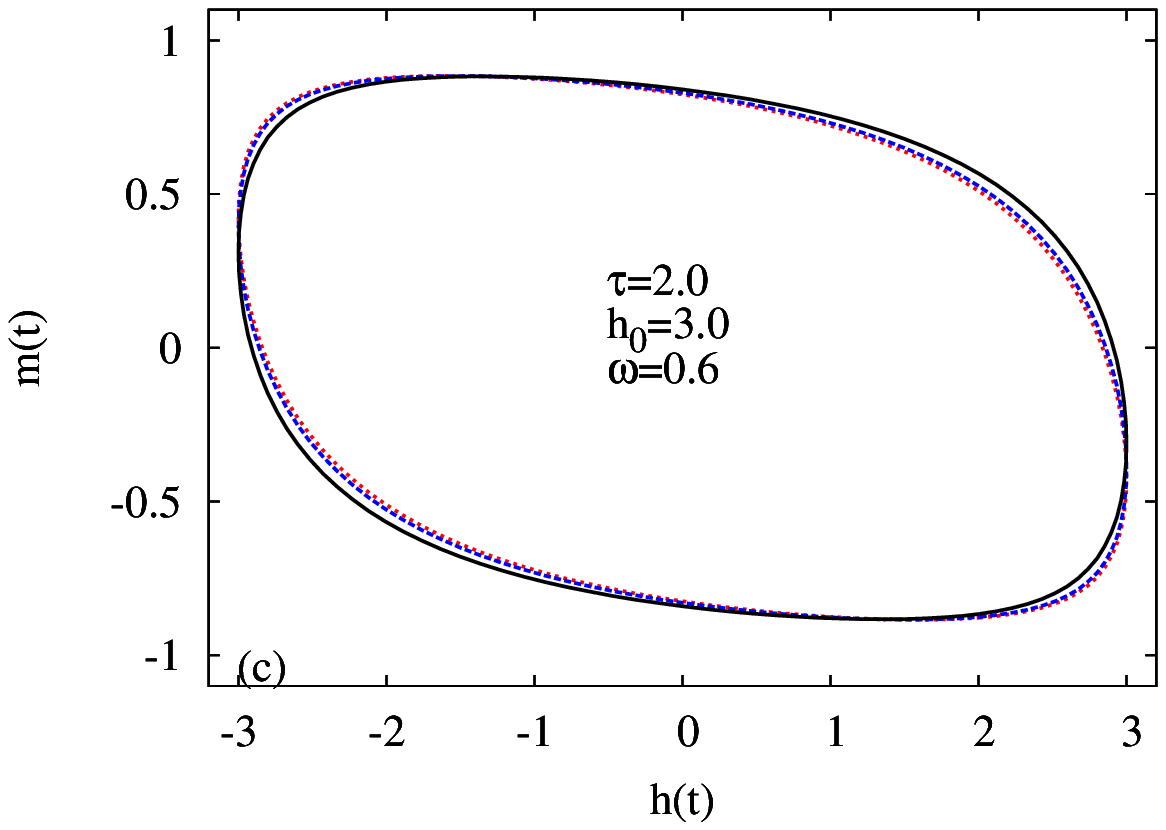, width=7.0cm}
\epsfig{file=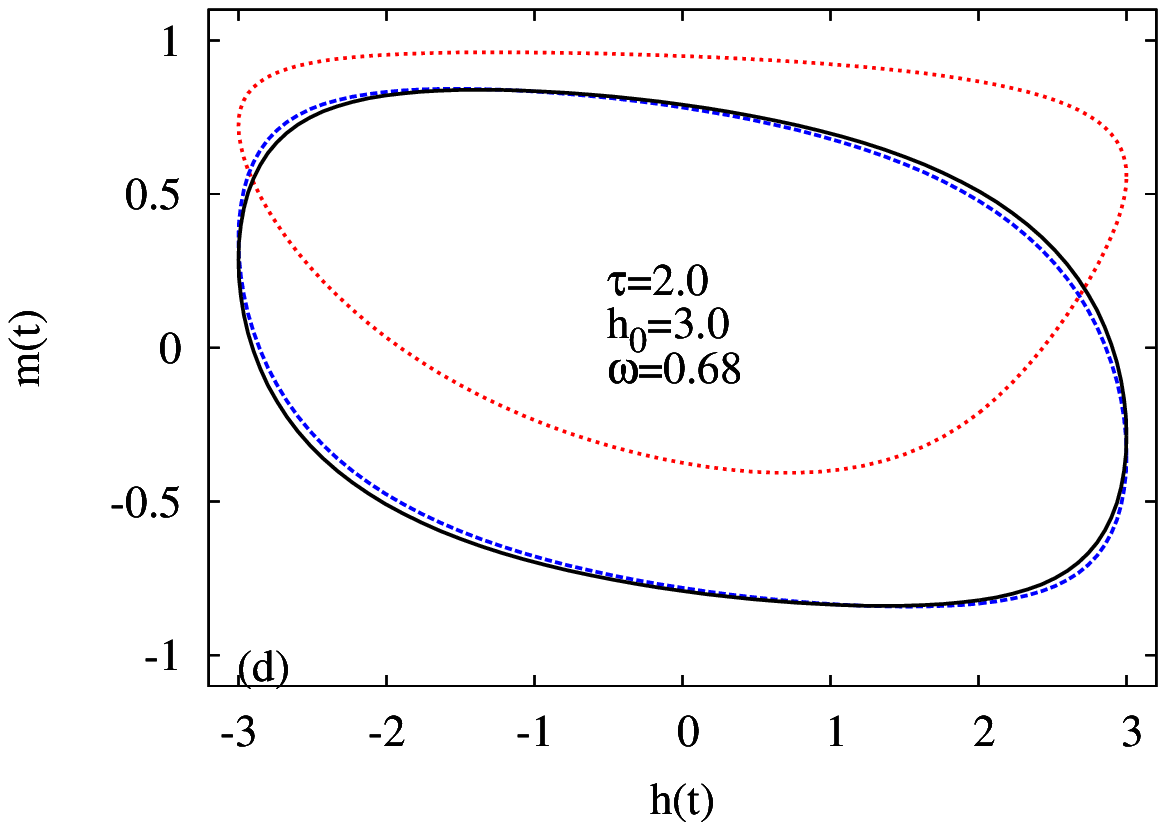, width=7.0cm}
\epsfig{file=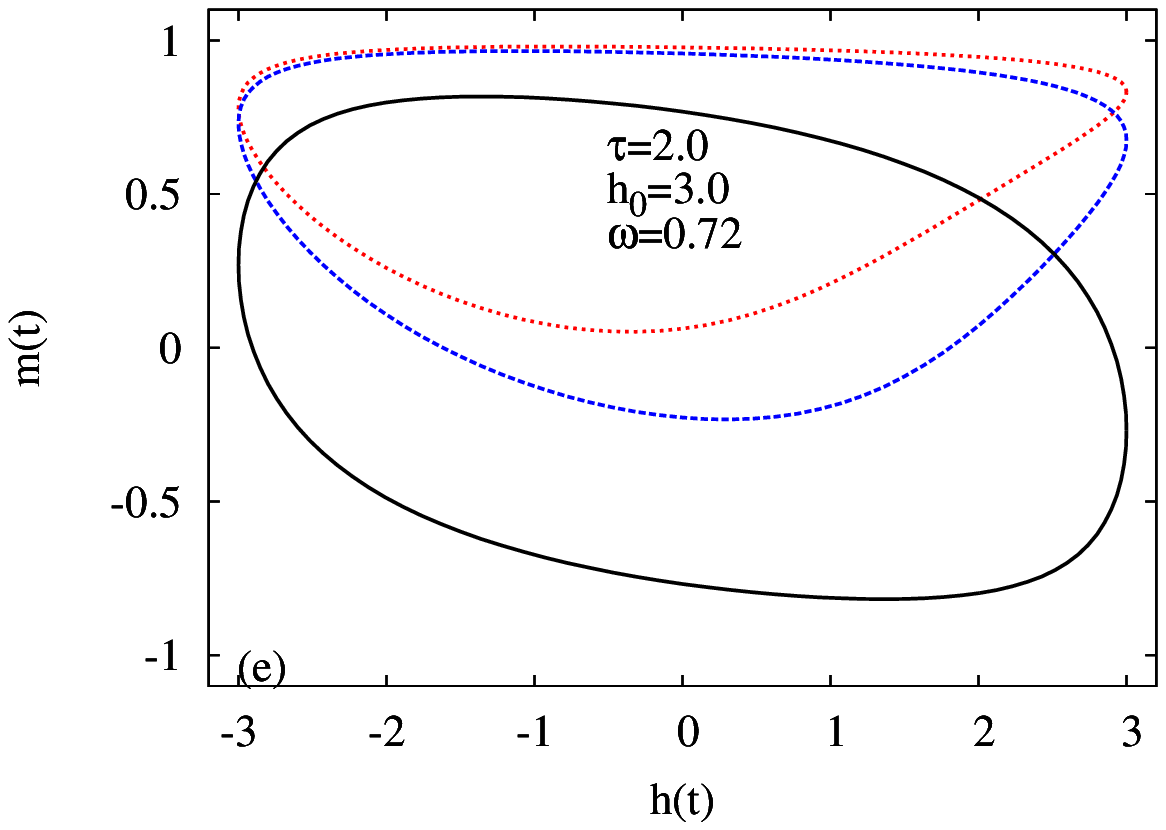, width=7.0cm}
\epsfig{file=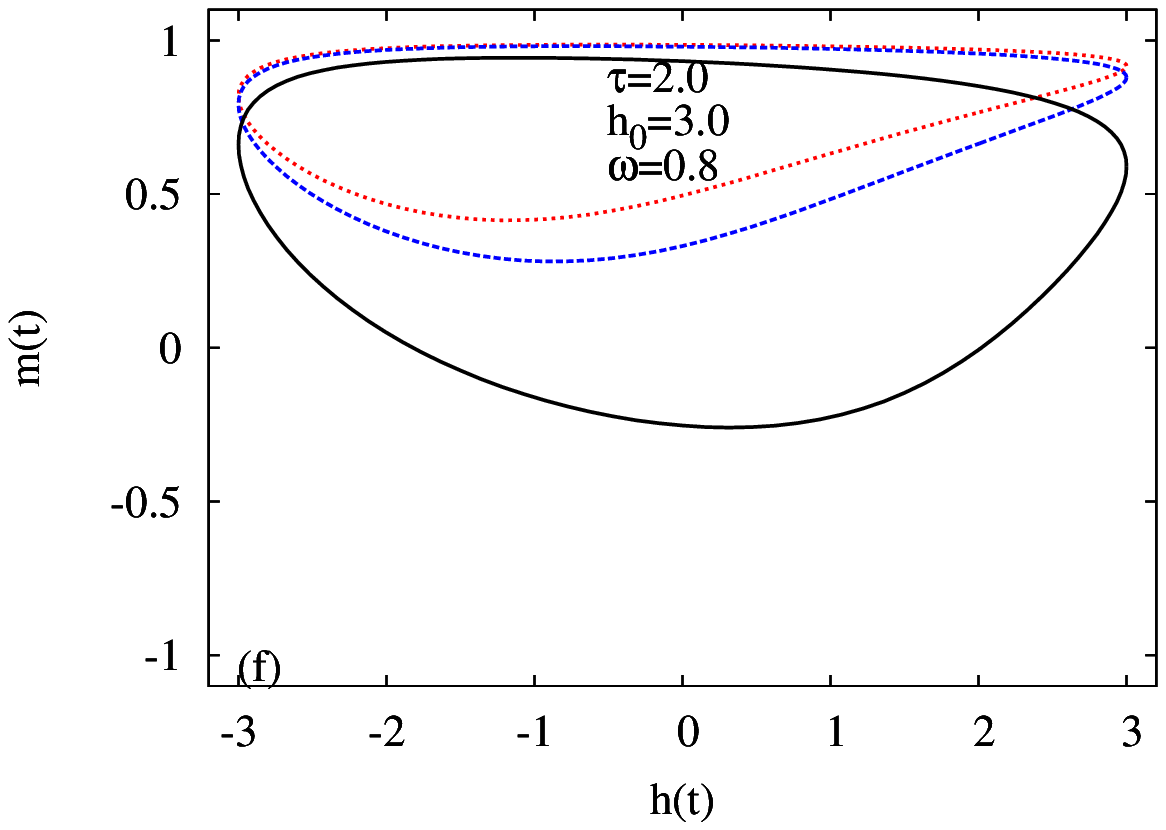, width=7.0cm}
\end{center}
\caption{Hysteresis loops for the  values of $\tau=2.0$ and $h_0=3.0$, with selected 
values of frequency.  The dotted  (red curve) corresponds to the Ising 
model ($\Delta=0.0$), dashed line (blue curve) corresponds to the XXZ 
type anisotropic Heisenberg model (with $\Delta=0.5$), while solid 
line (black curve) corresponds to the isotropic Heisenberg 
model ($\Delta=1.0$).} \label{sek9}\end{figure}

Lastly we want to touch upon the relation between the HLA values of Ising 
and isotropic Hesenberg model. We have concluded from Fig. \re{sek7} that,
greatness relationship of the HLA of these two model has been changed
while $\omega$ rises, for certain values of $h_0,\tau$. We have denoted 
this frequency value as $\omega^*$.  For lower frequencies, HLA of the 
Ising model is greater than that of the isotroic Heisenberg model while 
after the value of $\omega^*$, HLA of the isotropic Heisenberg model is 
greater than that of the Ising model. Now a question naturally arise: is 
this behavior general for all amplitude and temperature values? What is 
the dependence of the $\omega^*$ on other parameter values? We have to 
mention here that, there is no special reason to talk about this $\omega^*$ value. 
If we have held $\omega$ and $h_0$ constant, we would have talking about the $\tau^*$, 
which is defined in a similar way of $\omega^*$.

The answer of the first question is, mentioned behavior is not general for all amplitude 
and temperature values. Our calculations show that, in the ferromagnetic phase of the 
Ising model, HLA of the Ising model always little than the isotropic Heisenberg model. 
Of course not for very low temperature and amplitude values. For these values, 
hysteresis loop is nothing but almost a line which is parallel to the $h(t)$ axis.
Ising model is highly anisotropic, then spins tend to align in same direction 
along the anisotropy axis, in the ferromagnetic phase. Thus it is more difficult 
to magnetic field affect on this situation, in comparison with the isotropic Heisenberg model.
This can be seen also in Figs. \re{sek9} (e) and (f). Thus in the ferromagnetic 
phase of the Ising model, the effect of the rising anisotropy in the exchange 
interaction is simply decreasing of the HLA while other parameters are fixed. 
In other words, at a given ($h_0,\tau$) s, if the Ising model can show the DPT 
from the paramagnetic phase to the ferromagnetic phase, when the frequency rises, 
then   $\omega^*$ will appear at a certain value of the frequency. But if the Ising 
model cannot show the DPT, at that pair of  ($h_0,\tau$),  $\omega^*$ will not appear. 
If $\omega^*$ will not appear, the HLA of the Ising model  is always little than the 
isotropic Heisenberg model.

Now the second question, if $\omega^*$ present, then how it changes while the 
values of temperature  and amplitude change? The variation of the HLA with 
frequency of Ising and isotropic Heisenberg model for some choosen values 
of ($h_0,\tau$)  can be seen in Fig. \re{sek10}.
As seen in Fig. \re{sek10} that, for a fixed value of $h_0$, when the 
temperature rises,  $\omega^*$ increases (compare Fig. \re{sek10} (a) and (b)).  
On the other hand the effect of the rising $h_0$ at a fixed temperature can be 
seen in Figs. \re{sek10} (c), (a) and (d).  As seen in Figs. \re{sek10} (c), (a) 
and (d) rising $h_0$ causes to increase of $\omega^*$.

\begin{figure}[h]\begin{center}
\epsfig{file=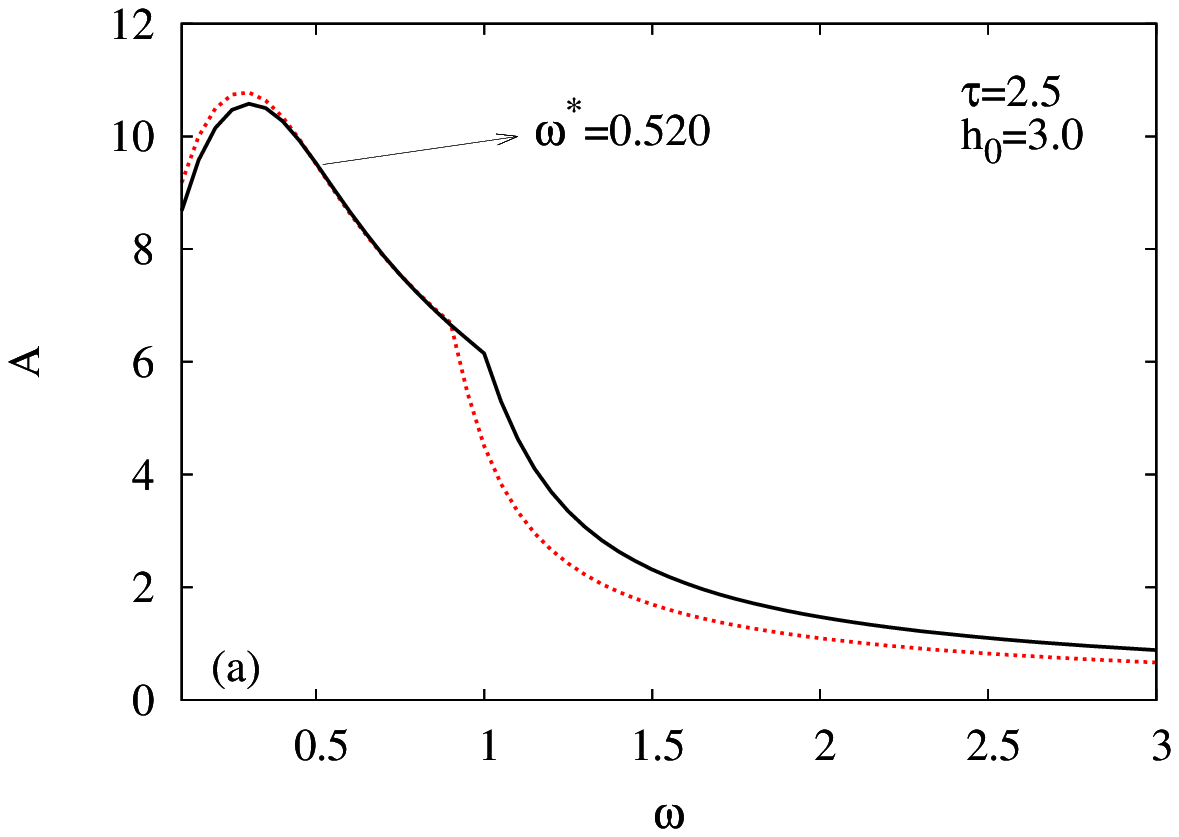, width=7.0cm}
\epsfig{file=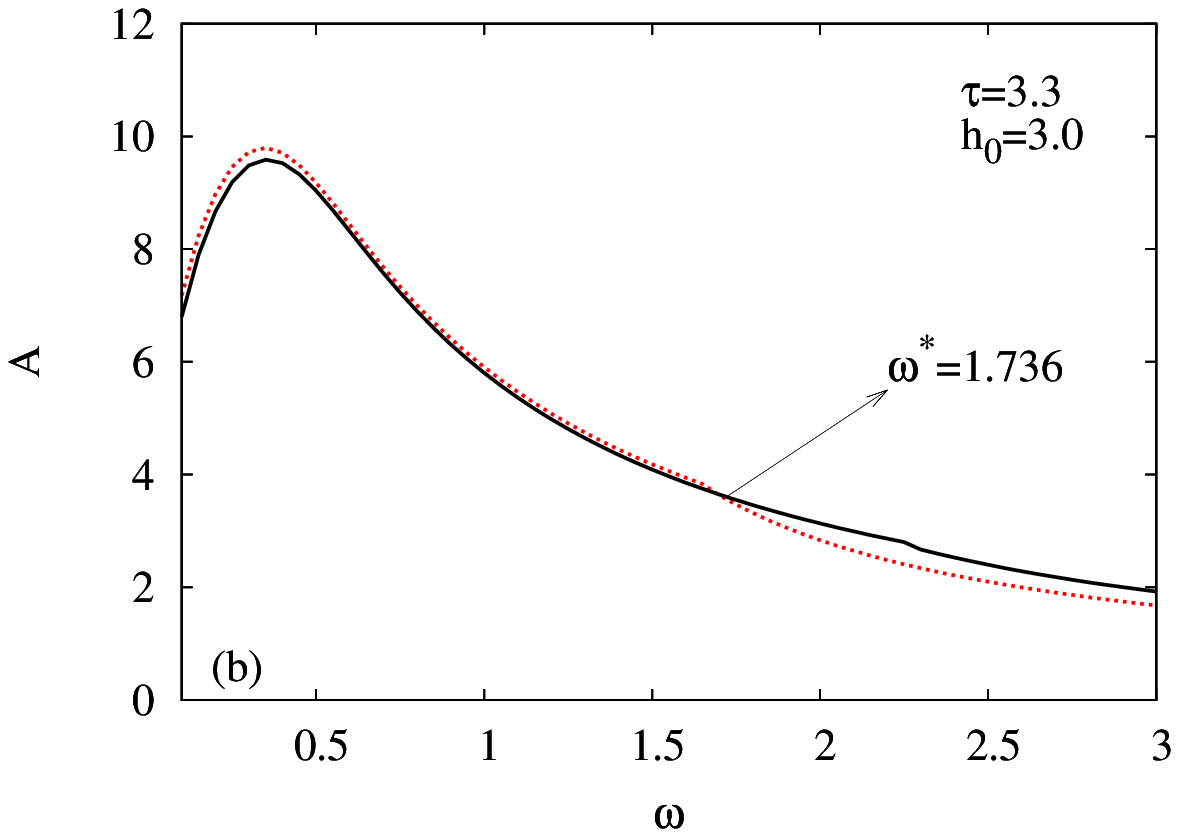, width=7.0cm}
\epsfig{file=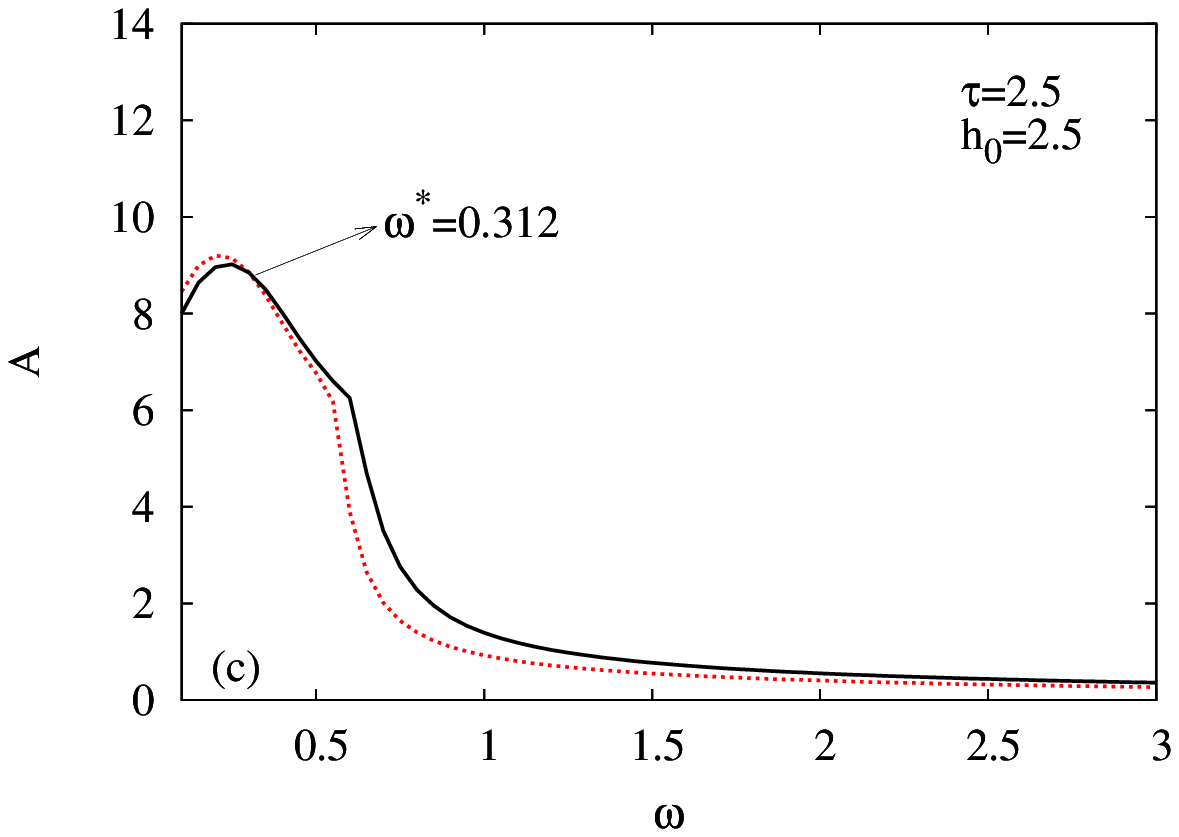, width=7.0cm}
\epsfig{file=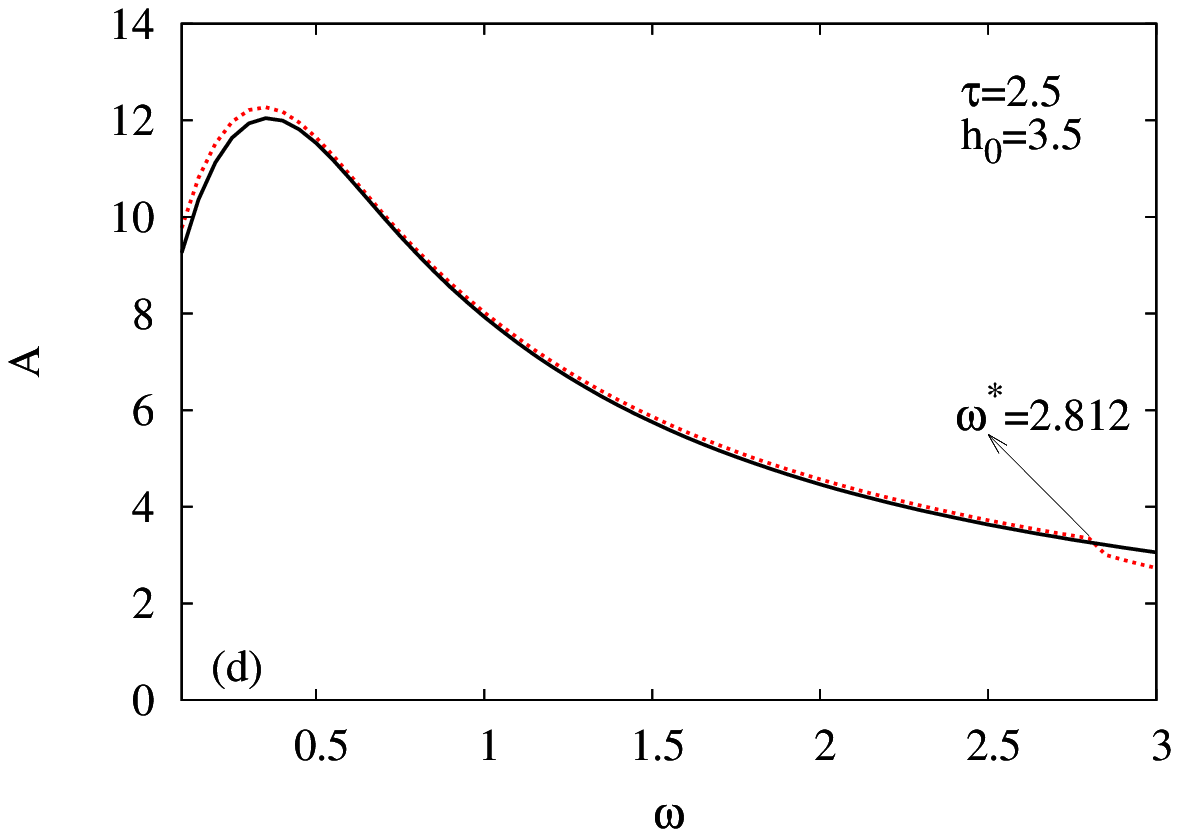, width=7.0cm}
\end{center}
\caption{Variation of the HLA with frequency of the magnetic field, for the 
selected values of ($h_0,\tau$) pairs as (a) $(3.0,2.5)$, (b) $(3.0,3.3)$, (c) $(2.5,2.5)$ and (d) $(3.5,2.5)$.  
The dotted line (red curve) corresponds to the Ising model ($\Delta=0.0$),  
while solid line (black curve) corresponds to the isotropic Heisenberg model 
($\Delta=1.0$). The value of the intersection frequency ($\omega^*$) of the 
two mentioned model is shown for each ($h_0,\tau$) pairs.} \label{sek10}\end{figure}

In conclusion, both of rising  $h_0$ and $\tau$ (while the other one is fixed) causes to 
increase of $\omega^*$, if it is present. We can see this in Fig. \re{sek11}, which is 
the variation of the $\omega^*$ with $\tau$ (shown by dashed lines) for selected 
values of $h_0=2.0,2.5,3.0,3.5$, in comparison with the $\omega_c$  of the Ising 
model (shown by solid lines). For the point $(\omega,\tau)$  which is under 
the $h_0$ (solid) curve, we say that Ising model is in the paramagnetic phase 
at that parameter values of $h_0,\omega,\tau$. In a same way,  for the 
point $(\omega,\tau)$ which is under the $h_0$ (dashed) curve  we say that 
HLA of the Ising model is greater than the isotropic Heisenberg model at 
that $h_0,\omega,\tau$ values. The interesting situation is that, not
for all of the paramagnetic phase (of the Ising model), HLA of the Ising model 
is greater than that of the isotropic Heisenberg model (compare solid and 
dashed lines related to any of the $h_0$ value).  This can be seen in in 
Fig. \re{sek11} as, difference between the  $\omega_c$ and $\omega^*$ 
curves in the lower values of the tempearture, for any of the 
chosen $h_0$. In contrast to this, for higher values of the 
temperature the value of $\omega^*$ getting same as the $\omega_c$.

\begin{figure}[h]\begin{center}
\epsfig{file=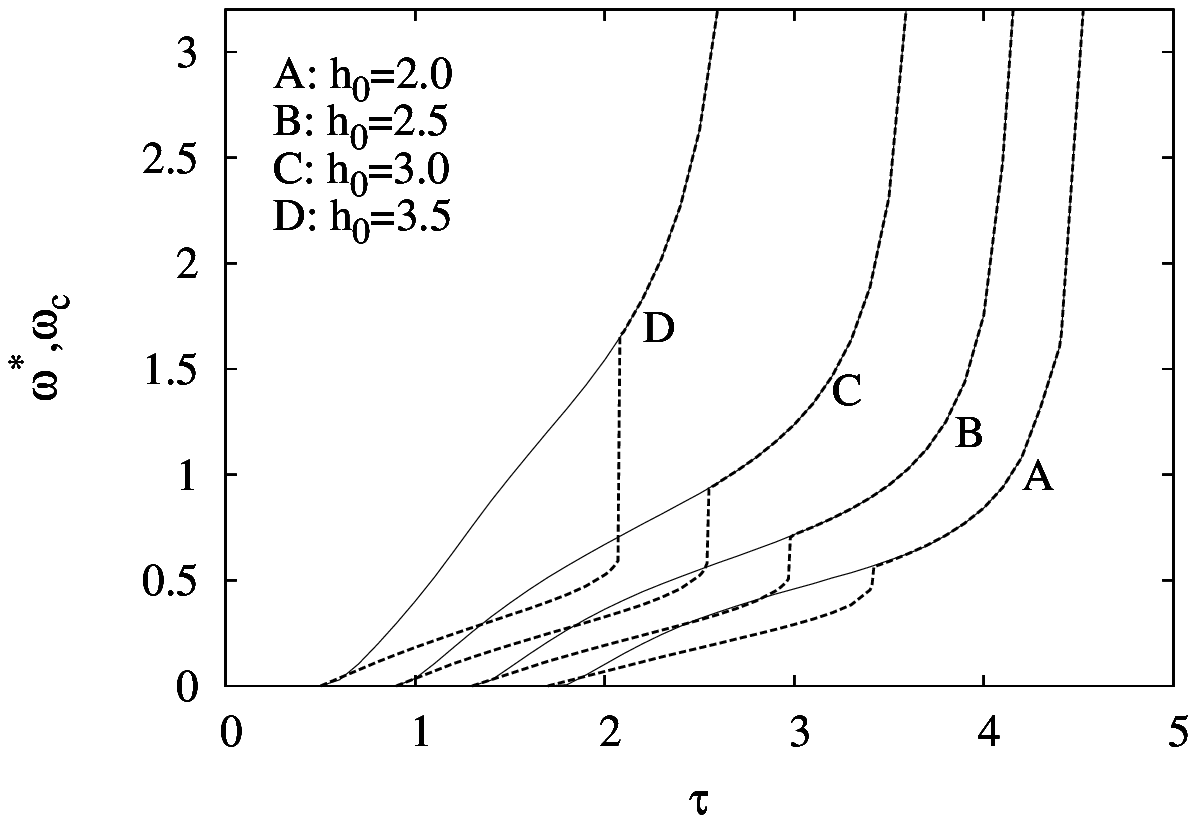, width=10.0cm}
\end{center}
\caption{Variation of the $\omega^*$  with the temperature for selected
values of the amplitude of the magnetic field $h_0$. Solid lines correspond to the  
critical frequency values $\omega_c$ of the Ising model, while  $\omega^*$ 
values shown by  dotted lines.} \label{sek11}\end{figure}

\section{Conclusion}\label{conclusion}

In conclusion, for a simple cubic lattice, the effect of the anisotropy 
in the exchange interaction on the dynamic phase diagrams and hysteresis 
loops within the anisotropic quantum Heisenberg model driving 
by sinusoidal time dependent magnetic field has been
investigated by benefiting from  effective field theory for two spin cluster. 
Dynamics of the system is defined by using Glauber-type stochastic process. 

First of all, the dynamic phase diagrams of the simple cubic lattice have been drawn in the $(h_{0}-\tau_c)$ 
plane for two limits of the model namely, isotropic Heisenberg model and its highly anisotropic 
limit, Ising model. It is concluded that, rising anisotropy in the exchange interaction causes to
increase of the critical temperature as well as critical field amplitude. On the other hand, 
rising anisotropy in the exchange interaction causes to decline of the critical frequency value.  
Besides, dynamic first order transitions appears at higher magnetic field amplitude and lower 
temperature values at all of the anisotropy in the exchange interaction values.

Instead of plotting the hysteresis loops for different possible values of the
anisotropy values, they have been treated based on three fundamental properties HLA, 
CF and RM. One advantage of the EFT is capability of the investigating large ranges 
of Hamiltonian parameters due to the short computation time in contrast to (quantum) MC. 
After rewieving the general effect of the rising frequency on the   HLA, CF and RM both 
for isotropic Heisenberg and Ising model, detailed investigation devoted on the effect 
of the rising anisotropy in the exchange interaction on the HLA. It is concluded that 
for the ferromagnetic phase of the Ising model, HLA of the Ising model is lower than 
the isotropic Heisenberg model within the same values of the parameters, 
namely $\omega,\tau,h_0$. Rising anisotropy in the exchange interaction on 
the HLA in this region causes decreasing HLA.  But, when the Ising model is 
in the paramagnetic phase, this situation does not hold. After a specific value 
of the frequency (which is denoted by $\omega^*$), HLA of the Ising model is 
little than the isotropic Heisenberg model. The variation of the $\omega^*$ with   
$\tau$ is given for different values of $h_0$. Then, for the paramagnetic phase 
of the Ising model, EFT-2 formulation gives the conclusion: for the values that 
provide $\omega<\omega^*$  rising anisotropy in the exchange interaction causes 
to increase of the HLA, at a fixed values of amplitude and temperature.
 Of course, this $\omega^*$ value depends on the amplitude and temperature.  
 This point   needs to be verified more accurate formulations such as (quantum) MC.

We hope that the results  obtained in this work may be beneficial 
form both theoretical and experimental point of view.

\end{document}